\documentclass{cta-author}
\usepackage{CJKutf8}
\usepackage{amsmath,amssymb,amsfonts}
\usepackage{algorithmic}
\usepackage{xcolor}
\usepackage{booktabs}
\usepackage{graphicx}
\usepackage[justification=centering]{caption}
\usepackage{enumitem}
{}
{}
{}
{}
{}
\usepackage{algorithm}
\usepackage{flushend}
\usepackage{setspace}
\usepackage{graphicx}
\usepackage{float}
\usepackage{subfig}
\usepackage{multirow}
\usepackage{makecell}
\usepackage{array}
\usepackage{booktabs}
\usepackage{url}
\begin{document}
\supertitle{Research Article}

\title{UAV Trajectory and Bandwidth Allocation for Efficient Data Collection in Low-Altitude Intelligent IoT: A Hierarchical DRL Approach}

\author{\au{Zhenjia Xu$^{1}$}, \au{Xiaoling Zhang$^{2}$}, \au{Nan Qi$^{1\corr}$}, \au{Guangxu Zhu$^{3}$}, \au{Xiaojie Li$^{4}$}, \au{Luliang Jia$^{5}$}}

\address{\add{1}{Key Laboratory of Dynamic Cognitive System of Electromagnetic Spectrum Space, Ministry of Industry and Information Technology, Nanjing University of Aeronautics and Astronautics, Nanjing, China, 210016.}
\add{2}{School of Artificial Intelligence, Shenyang University of Technology, Shenyang 110870, China}
\add{3}{Shenzhen Research Institute of Big Data, Shenzhen, 518172, China}
\add{4}{National Mobile Communication Research Laboratory, Southeast University, Nanjing 210096, China}
\add{5}{School of Space Information, Space Engineering University, Beijing 101416, China}
\email{xuzhenjia1014@gmail.com and nanqi.commun@gmail.com}}

\begin{abstract}
The low-altitude Internet of Things (IoT), supported by unmanned aerial vehicles (UAVs), provides ground sensing networks with advanced real-time monitoring and data collection. To maximize data collection volume from distributed IoT nodes, AI-powered data collection technology plays a critical role in enabling intelligent decision-making. Among them, deep reinforcement learning (DRL) has gained particular attention. However, existing DRL-based work on UAV-assisted IoT data collection rarely addresses challenges such as interference and dynamic data volume, while also suffering from high computational demands and slow convergence. To address these challenges, a hierarchical DRL (HDRL) is designed to optimize UAV trajectories and bandwidth allocation to maximize data collection volume. Firstly, the proposed scenario incorporates interference, dynamic data volume of IoT nodes, and multiple types of obstacles. The entire task is hierarchically structured: the upper-level makes flight trajectory decisions at a coarse temporal granularity, while the lower-level makes bandwidth allocation decisions at a finer temporal granularity. Secondly, a trajectory and bandwidth allocation optimization algorithm based on hierarchical deep deterministic policy gradients (TBH-DDPG) is proposed to solve the problem. Finally, simulation results demonstrate that the proposed algorithm improves convergence speed by $44.44\%$, and reduces computational cost by $58.05\%$, compared to non-hierarchical algorithm.
\end{abstract}

\maketitle

\section{Introduction}
\subsection{Background and motivation}
Against the backdrop of 6G wireless network evolution, the low-altitude intelligent Internet of Things (IoT) is emerging as a critical infrastructure for supporting smart cities and next-generation IoT services \cite{11169508}. Unmanned aerial vehicles (UAVs) not only serve as traditional communication relays but also play an active role in environmental sensing and real-time intelligent decision-making \cite{mu2023uav}. Within this framework, UAVs collaborate with terrestrial IoT nodes to construct a dynamic, context-aware sensing and communication network \cite{11268973}. This network enables UAVs to efficiently perform large-scale data collection tasks such as urban situational awareness \cite{dehghan2024review}, traffic flow analysis \cite{butilua2022urban}, and environmental pollution monitoring \cite{motlagh2023unmanned}, laying a solid foundation for the development of autonomous systems and the advancement of smart cities.

UAVs are characterized by high mobility, ease of deployment, and low cost \cite{valavanis2014handbook}. IoT nodes can monitor and collect environmental data. Consequently, UAVs acting as aerial mobile base stations can then gather the data sensed by these IoT nodes, significantly reducing labor costs and enhancing operational efficiency \cite{mozaffari2017mobile}. Compared to traditional ground-based communication systems, UAVs are more likely to have line-of-sight (LoS) air-to-ground channels \cite{lu2024joint} and closely approach distributed IoT nodes, thereby providing higher communication quality and effectively reducing the transmission energy consumption of IoT nodes \cite{wang2021trajectory}. In addition, many agricultural IoT nodes are located in remote, mountainous, or areas where internet access costs are relatively high. The use of UAVs can effectively reduce operational and maintenance costs in such scenarios \cite{mingcheng2023deep}. However, data collection by UAVs encounters significant challenges, including power constraints, obstacles, and interference sources \cite{fu2021energy}. Therefore, ensuring that UAVs collect the maximum data volume in limited timeframes while completing tasks efficiently and reliably has remained a critical and valuable research problem.

Significant progress has been made in the field of UAV-assisted data collection for IoT nodes. Due to the highly non-convex nature of optimization problems in this domain, solving them has been considerably challenging. Traditional approaches have employed convex optimization methods \cite{feng2021uav, zhan2017energy, chen2023minimizing} to reformulate these problems for solution. However, these methods come with several limitations, such as a substantial increase in problem complexity after transformation, difficulties in balancing multi-objective requirements, and poor adaptability to dynamic environments. Heuristic algorithms \cite{10639960}, such as genetic algorithms (GA) \cite{liu2018age} and monarch butterfly optimization algorithm (MBO) \cite{zhou2021uav}, have also been used to address these problems. Nevertheless, these algorithms face high computational complexity when applied to high-dimensional, multi-constrained problems \cite{huang2022simulated}. Additionally, in dynamic environments, their ability to adjust in real-time is limited, making it challenging to adapt to environmental changes and prone to getting trapped in local optima.

In recent years, researchers have increasingly explored the use of deep reinforcement learning (DRL) algorithms \cite{mnih2013playing} for UAV data collection. A summary and comparison of the relevant literature are presented in Table \ref{tab1}. Unlike traditional methods, DRL typically models the problem as a Markov Decision Process (MDP) \cite{sutton2018reinforcement}, where the UAV acts as an agent, while the states of IoT nodes, UAV states, and other factors constitute the environment. The objective function is transformed into rewards received by the agent from the environment based on its actions. The algorithm aims to maximize cumulative rewards, allowing the agent to iteratively learn through interactions with the environment. As a result, it becomes possible to effectively solve problems without focusing on the mathematical modeling of highly non-convex optimization problems. Existing studies have focused mainly on optimization of UAV trajectory \cite{bayerlein2020uav,  wang2024trajectory, fan2022ris}, joint optimization of UAV trajectory and bandwidth allocation \cite{ding20203d, luo2020intelligent}, optimization of energy consumption for UAVs and IoT nodes \cite{zhu2021uav, liu2019energy}, and optimization of UAV trajectory and the age of information (AoI) for collected IoT data \cite{sun2021aoi, yi2020deep}. \cite{9762906} introduced a CNN-DQN-based resource scheduling scheme for UAV-assisted emergency communication networks, while \cite{9892691} proposed a Transformer-A$^\ast$-based UAV trajectory planning method for AoI-minimal data collection. However, applying DRL methods to real-world scenarios presents significant challenges, including high requirements for training data, high-dimensional action-state spaces, and sparse rewards \cite{dulac2019challenges}. \cite{du2024distributed} introduces transfer learning to accelerate the QMix algorithm for multi-UAV data collection, which significantly speeds convergence. However, the model’s computational burden remains substantial. Efficiently managing tasks, accelerating network convergence, and reducing training parameters to accommodate the limited computational capabilities of UAVs are critical concerns \cite{he2023uav, qu2023elastic}. Increasing attention has been directed to hierarchical deep reinforcement learning (HDRL) to address these issues. \cite{dietterich2000hierarchical}.

HDRL has emerged as an approach that decomposes complex decision-making tasks into several subtasks, using a hierarchical structure. The higher level operates on a coarse temporal granularity to perform global task planning, while the lower-level policy focuses on executing specific actions to complete subtasks at a finer temporal granularity. By breaking down objectives and utilizing multi-level rewards, HDRL effectively reduces the dimensionality of the state-action space, enabling more efficient learning and decision-making. Additionally, HDRL has addressed the mismatch between temporal and action scales in agents \cite{kulkarni2016hierarchical}. For instance, when an agent executes various actions with differing durations, temporal and action-level hierarchies effectively avoid the exponential growth of output dimensions in joint action networks, which would otherwise increase the difficulty of training. In the context of UAV-assisted IoT node data collection, existing studies have primarily designed HDRL algorithms based on the option framework \cite{xu2022optimization, qin2022uav, zhang2020hierarchical, zhiyu2020research}. These problems were typically modeled as semi-markov decision processes (SMDP), extending MDPs by introducing higher-level action options to guide the lower-level action space. This design has also allowed for flexible control over subtask durations. Simulation results have shown promising improvements in performance and convergence. However, in existing SMDP models, the higher-level options are often directly associated with predefined sequences of lower-level actions. In other words, once an action option is selected, the corresponding lower-level actions are fixed. This constrains the agent’s exploration space and impairs the model’s generalization capabilities. By contrast, the proposed SMDP model introduces a novel decision-making layer for bandwidth allocation within lower-level actions after an option is selected. Overall, while HDRL applications in UAV-assisted data collection have remained relatively scarce, they have demonstrated significant potential in accelerating network training, reducing training difficulty, and minimizing network size. 

Furthermore, existing system scenarios in this field are often heavily simplified and constrained. For example, it is typically assumed that UAVs can only collect data directly above IoT nodes, that the environmental scenes contain a small number of obstacles, or that the power and positions of jammers within the scene are known \cite{qi2022coalitional, wang2022energy}. While these restrictive assumptions have facilitated theoretical analysis, they have also limited the practical applicability of research outcomes in real-world scenarios.

\begin{table*}[!t]
	\caption{Related Literature Organization and Comparison}
	\label{tab1}
	\centering
	\scriptsize
	\setlength{\tabcolsep}{2.5pt} 
	\renewcommand{\arraystretch}{1.2} 
	\begin{tabular}{@{}|l|l|l|l|l|l|l|l|@{}}
		\toprule
		Methods & Reference & Obstacles exist & Hierarchical & Bandwidth allocation & Algorithm & Number of UAVs & Optimization objective \\ \midrule
		\multirow{3}{*}{\begin{tabular}[c]{@{}l@{}}Convex \\ Optimization\end{tabular}} 
		& \cite{feng2021uav} & × & × & × & AO & Single & Maximize average data rate throughput \\ \cmidrule(l){2-8} 
		& \cite{zhan2017energy} & × & × & × & SCO & Single & Minimize the maximum energy consumption of all IoT nodes \\ \cmidrule(l){2-8} 
		& \cite{chen2023minimizing} & × & × & × & CVX & Single & Minimize the age of collected data \\ \midrule
		
		\multirow{3}{*}{\begin{tabular}[c]{@{}l@{}}Heuristic \\ Algorithms\end{tabular}} 
		& \cite{liu2018age} & × & × & × & GA & Single & Minimize the age of collected data \\ \cmidrule(l){2-8} 
		& \cite{zhou2021uav} & × & × & × & MBO & Single & Maximize data collection efficiency \\ \cmidrule(l){2-8} 
		& \cite{huang2022simulated} & × & × & × & SA-PSO & Single & Minimize the time of data collection \\ \midrule
		
		\multirow{14}{*}{\begin{tabular}[c]{@{}l@{}}DRL \\ Algorithms\end{tabular}} 
		& \cite{bayerlein2020uav} & \checkmark & × & × & DDQN & Single & Maximize the amount of data collected \\ \cmidrule(l){2-8} 
		& \cite{wang2024trajectory} & \checkmark & × & × & DDQN & Single & Maximize the amount of data collected \\ \cmidrule(l){2-8} 
		& \cite{fan2022ris} & × & × & × & SAC & Single & Minimize the age of collected data \\ \cmidrule(l){2-8} 
		& \cite{ding20203d} & × & × & \checkmark & DDPG & Single & Maximizing fairness in data collection \\ \cmidrule(l){2-8} 
		& \cite{luo2020intelligent} & × & × & \checkmark & DQN & Single & Maximize the completion rate of data collection \\ \cmidrule(l){2-8} 
		& \cite{zhu2021uav} & × & × & × & A2C & Single & Maximize data collection efficiency \\ \cmidrule(l){2-8} 
		& \cite{liu2019energy} & \checkmark & × & × & MADDPG & Multiple & Maximize data collection efficiency \\ \cmidrule(l){2-8} 
		& \cite{sun2021aoi} & × & × & \checkmark & TD3 & Single & Minimize the age of collected data \\ \cmidrule(l){2-8} 
		& \cite{9762906} & × & × & \checkmark & CNN-DQN & Multiple & Maximize spectrum efficiency \\ \cmidrule(l){2-8}
		& \cite{9892691} & × & × & × & Transformer-A$^\ast$ & Single & Minimize the age of collected data \\ \cmidrule(l){2-8}
		& \cite{du2024distributed} & \checkmark & × & \checkmark & QMix & Multiple & Maximizing fairness in data collection \\ \cmidrule(l){2-8} 
		& \cite{xu2022optimization} & × & \checkmark & × & CT-MADDPG & Multiple & Maximize the amount of data collected \\ \cmidrule(l){2-8} 
		& \cite{qin2022uav} & × & \checkmark & × & H-DQN & Single & Minimize the age of collected data \\ \cmidrule(l){2-8} 
		& \cite{zhang2020hierarchical} & × & \checkmark & × & MADOL & Multiple & Minimize the time of data collection \\ \cmidrule(l){2-8} 
		& \cite{zhiyu2020research} & × & \checkmark & × & H-DQN & Single & Minimize the time of data collection \\ \cmidrule(l){2-8} 
		& our work & \checkmark & \checkmark & \checkmark & TBH-DDPG & Single & Maximize the amount of data collected \\ \bottomrule
	\end{tabular}
\end{table*}

\subsection{Contributions}
We considered a practical system environment that includes multiple obstacle regions, interference, and dynamic data volume of IoT nodes. The focus has been on optimizing UAV trajectories and bandwidth allocation to maximize data collection volume from IoT nodes. A trajectory and bandwidth allocation optimization algorithm based on hierarchical deep deterministic policy gradients (TBH-DDPG) has been proposed to solve the optimization problem. By hierarchically structuring the UAV trajectory planning strategy and the bandwidth allocation strategy, the proposed algorithm has demonstrated superior performance in terms of network computational load and convergence speed. The main contributions of our work are as follows.

\begin{itemize}[leftmargin=15pt,labelsep=6pt]

\item Our work addresses the data collection challenges of IoT nodes in complex scenarios by considering factors such as the real-time growth of IoT node data, communication obstacle areas on the ground, no-fly zones, and various interferences. 

\item The joint optimization of UAV flight trajectories and bandwidth allocation has been conducted to maximize the data collection volume from IoT nodes. Considering the multiple communications involved when the UAV performs flight actions, this has been transformed into an SMDP model, where the upper-level strategy focus on planning the UAV trajectory, and the lower-level strategy is used to plan the UAV’s bandwidth allocation.
 
\item A hierarchical deep reinforcement learning algorithm, TBH-DDPG, has been proposed. First, the UAV, as the agent, executes the upper-level flight strategy. Then, based on this flight strategy, the UAV executes the lower-level bandwidth allocation strategy, effectively reducing the state-action dimension.

\item In the simulation experiments, a comprehensive comparison of baseline algorithms was carried out in terms of convergence speed, performance in various scenarios, and the impact of the IoT node count and data volume growth rate. The simulation results show that the proposed TBH-DDPG algorithm maintains superior performance while achieving faster convergence and reduced network computational complexity.

\end{itemize}

The notations and their physical meanings for the entire system model are given in Tables \ref{tab22} and \ref{tab2}. The remainder of this work is organized as follows. Section \ref{System model} provides an overview of the system model and the problem formulation. Section \uppercase\expandafter{\romannumeral3} introduces the establishment of the SMDP model and the algorithm design of TBH-DDPG. Section \uppercase\expandafter{\romannumeral4} presents the comprehensive experimental results. Finally, Section \uppercase\expandafter{\romannumeral5} concludes the work.

\begin{table}[]
	\caption{Notation List}\label{tab22}
	\centering
	\setlength{\tabcolsep}{0.4pt} 
	\renewcommand{\arraystretch}{1.2} 
	\begin{tabular}{|c|c|}
		\hline
		$p_i$ & The $i$-th IoT node location (m)\\ \hline
		$p_j$ & The $j$-th jammer location (m)\\ \hline
		$p_u$ & The UAV loaction (m)\\ \hline
		$v_u$ & UAV flight speed (m/s)\\ \hline
		${\delta }_{n}$ & The $n$-th flight period (s) \\ \hline
		${\delta }_{n,m}$ & The $m$-th communication slot in ${\delta }_{n}$ (s)\\ \hline
		$L_i$ & UAV-to-IoT path loss (dB) \\ \hline
		$g_i$ & UAV-to-IoT channel gain \\ \hline
		$L_j$ & UAV-to-jammer path loss (dB) \\ \hline
		$g_j$ & UAV-to-jammer channel gain \\ \hline
		$d_i$ & UAV-to-IoT distance (m) \\ \hline
		$G_m$ & Main lobe gain of the jammer \\ \hline
		${\theta }_{m}$ & Jammer beamwidth (rad) \\ \hline
		$G_s$ & Isotropic antenna gain of the jammer \\ \hline
		$G_J$ & Antenna gain of the jammer \\ \hline
		$d_{m,j}$ & \begin{tabular}[c]{@{}c@{}}The distance from the UAV to the center-\\ line of the j-th jammer main beam (m) \end{tabular} \\ \hline
		$C_i$ & \begin{tabular}[c]{@{}c@{}}The transmission rate between\\ the UAV and the i-th IoT node\end{tabular} (bit/s) \\ \hline
		$D_i^{col}$ & \begin{tabular}[c]{@{}c@{}}The volume of data collected by\\ the UAV from the i-th IoT node\end{tabular} (Mb) \\ \hline
		$D_i^{rem}$ & The remaining data volume of the i-th IoT node (Mb) \\ \hline
	\end{tabular}
\end{table}

\section{System model}
\label{System model}
\subsection{UAV-assisted Data Collection Scenario Model}
As shown in Fig. 1, we consider a UAV-enabled data collection scenario in a square urban area within a low-altitude IoT system. Three types of obstacle zones are defined as follows: (1) zones that affect communication only, such as low-rise structures over which UAVs can fly; (2) zones that affect flight only, such as densely populated ground areas or no-fly zones; and (3) zones that affect both flight and communication, such as areas characterized by clusters of high-rise buildings.
There are $I$ ground IoT nodes, with the position of the $i$-th IoT node represented by ${p}_{i}=\begin{Bmatrix} {x}_{i},{y}_{i},0\end{Bmatrix}$, where $i\in \begin{Bmatrix}1,2,\dots ,I\end{Bmatrix}$. Additionally,  ground jammers are included. The position of the $j$-th jammer is represented by ${p}_{j}=\begin{Bmatrix} {x}_{j},{y}_{j},0\end{Bmatrix}$, where $j\in \begin{Bmatrix}1,2,\dots ,J\end{Bmatrix}$. The UAV departs from a take-off zone, flies at a fixed altitude $H$, and aims to maximize data collection volume from the $I$ IoT nodes while avoiding obstacle zones and minimizing interference. The mission is completed when the UAV lands successfully in the landing zone. The total mission time is represented by $T$. It is divided into $N$ equal flight periods, with the $n$-th flight period represented by ${\delta }_{n}$, where $n\in \begin{Bmatrix}1,2,\dots ,N\end{Bmatrix}$. Consequently, The UAV’s position at the $n$-th flight period can be represented as ${p}_{u}({\delta }_{n})=\begin{Bmatrix} {x}_{u}({\delta }_{n}),{y}_{u}({\delta }_{n}),H\end{Bmatrix}$. Given the small division of flight periods, the UAV’s velocity within each time slot can be assumed constant. It is further assumed that the UAV’s speed during each flight period is either 0 or $v$, that is, $v_{u}({\delta }_{n})\in \begin{Bmatrix} 0,v\end{Bmatrix}$.

\begin{figure}[!t]
	\centering
	\includegraphics[width=3.5in]{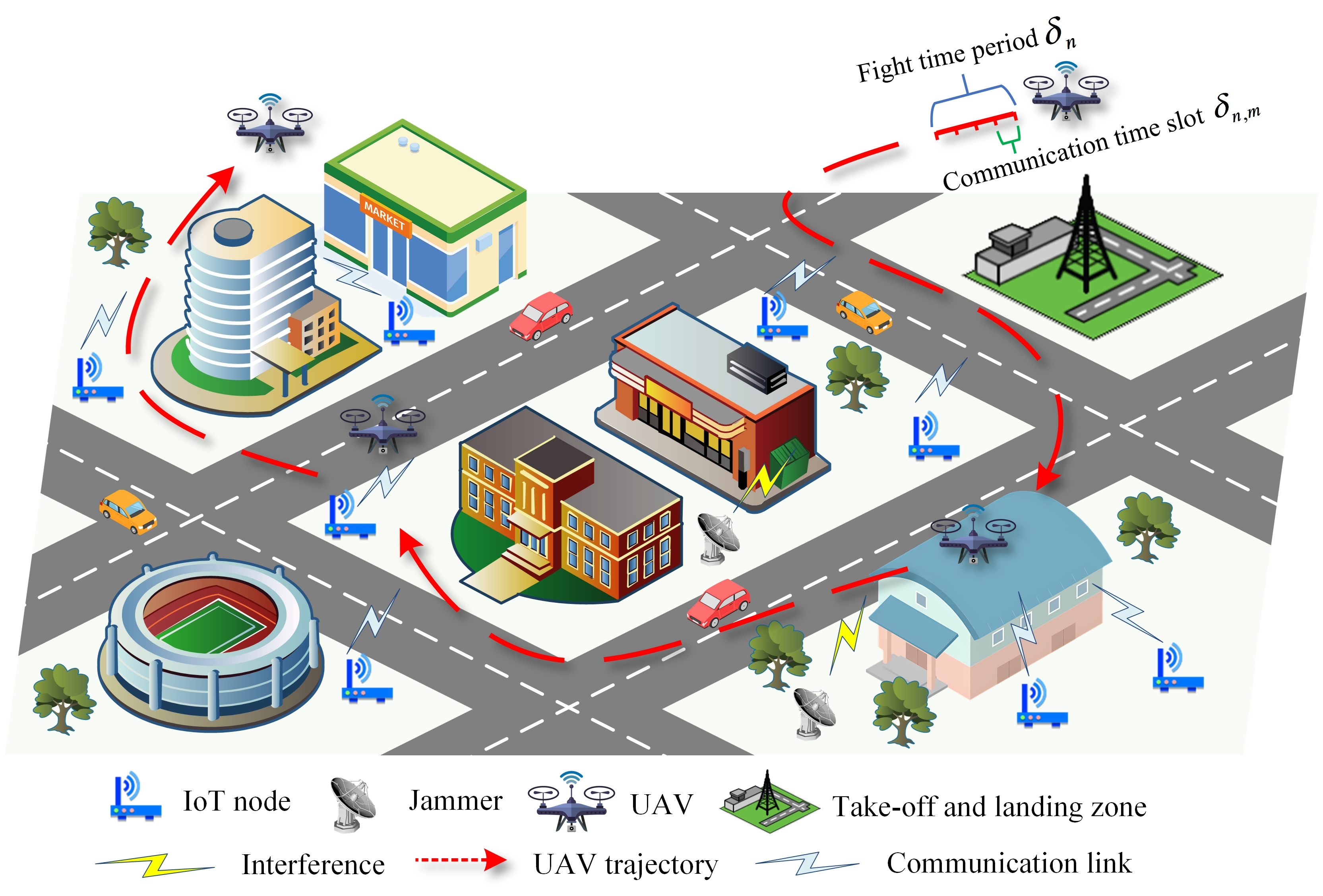}
	\caption{Data collection for the food processing industry in low-altitude IoT system scenario.}
	\label{fig1}
\end{figure}

\begin{figure}[!t]
	\centering
	\includegraphics[width=3in]{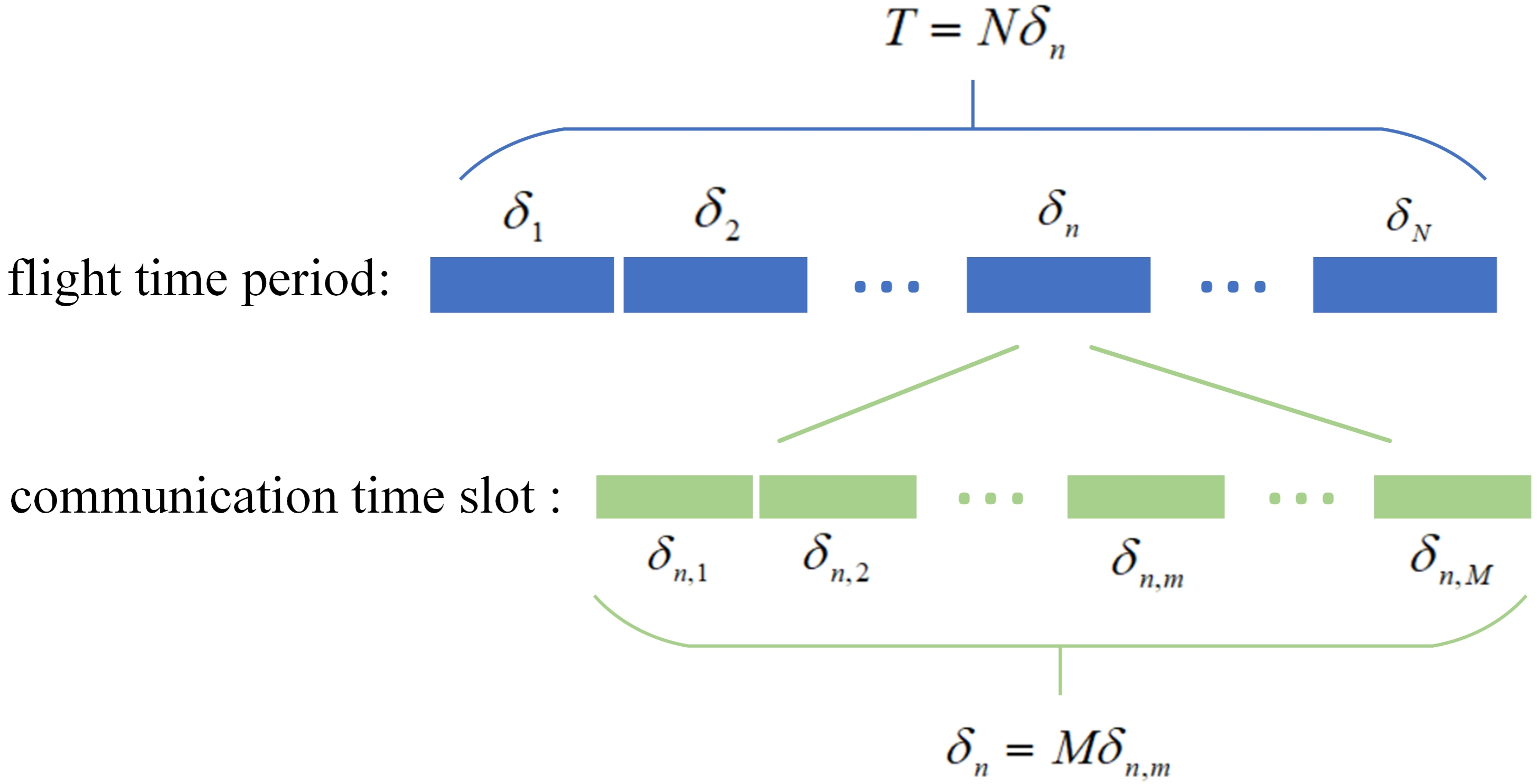}
	\caption{Time slot division.}
	\label{fig2}
\end{figure}

Considering that the UAV can perform multiple bandwidth allocation communications within a single flight period, each flight period ${\delta }_{n}$ is further divided into $M$ equally spaced communication time slots. The $m$-th communication time slot within the $n$-th flight period is represented by ${\delta }_{n,m}$, where $m\in \begin{Bmatrix}1,2,\dots ,M\end{Bmatrix}$. During any communication time slot, the UAV employs a frequency division multiple access (FDMA) scheme for communication. The slot division for the entire mission is shown in Fig. \ref{fig2}. Given the minimal position change of the UAV within each flight period, we simplify bandwidth allocation and communication by treating the UAV’s location as that at the beginning of the next flight period. For example, at communication time slot ${\delta }_{n,m}$, the UAV’s position is approximated by ${p}_{u}({\delta }_{n+1})$. Moreover, since the IoT nodes continuously collect data from the environment, their cumulative data volume increases over time. Without loss of generality, we assume that the accumulated data volume of each IoT node grows at a constant rate. After each communication time slot ${\delta }_{n,m}$, every IoT node's data volume increases by a fixed amount ${d}_{inc}$. However, the data storage capacity of each IoT node is limited, with the maximum capacity represented by ${D}_{max}$.

\subsection{LoS/NLoS Channel Communication Model}
Line-of-sight/non-line-of-sight (LoS/NLoS) communication channel model is considered. \cite{esrafilian2018learning}. This model accounts for the impact of shadowing and logarithmic fading on the channel. Moreover, there are multiple jammers. Specifically, during a communication time slot ${\delta }_{n,m}$, the dB-scale path loss $L_i({\delta }_{n,m})$ between the UAV and $i$-th IoT node is expressed as
\begin{equation}
	\label{eq1}
        {L}_{i}({\delta }_{n,m})=\begin{cases}{\alpha }_{LoS}\log_{10}{{d}_{i}({\delta }_{n,m})}+{\eta }_{LoS}&,if\; LoS\\ {\alpha }_{NLoS}\log_{10}{{d}_{i}({\delta }_{n,m})}+{\eta }_{NLoS}&,if\; NLoS\end{cases},
\end{equation}
where ${d}_{i}({\delta }_{n,m})$ represents the distance from the UAV to the 
$i$-th IoT node during time slot ${\delta }_{n,m}$, ${\alpha }_{LoS}$ and ${\alpha }_{NLoS}$ represent the path loss for $LoS$ and $NLoS$ channels, respectively. Similarly, ${\eta }_{LoS}$ and ${\eta }_{LoS}$ represent the effect of shadow fading for $LoS$ and $NLoS$ channels, which are modeled as ${\eta }_{l}\sim  \mathcal{N}(0,{\sigma}_{LoS}^{2})$ and ${\eta }_{l}\sim  \mathcal{N}(0,{\sigma}_{NLoS} ^{2})$, respectively. Accordingly, the corresponding linear channel power gain is given by $h_i(\delta_{n,m})=10^{-L_i(\delta_{n,m})/10}$.

\begin{figure}[!t]
	\centering
	\includegraphics[width=2in]{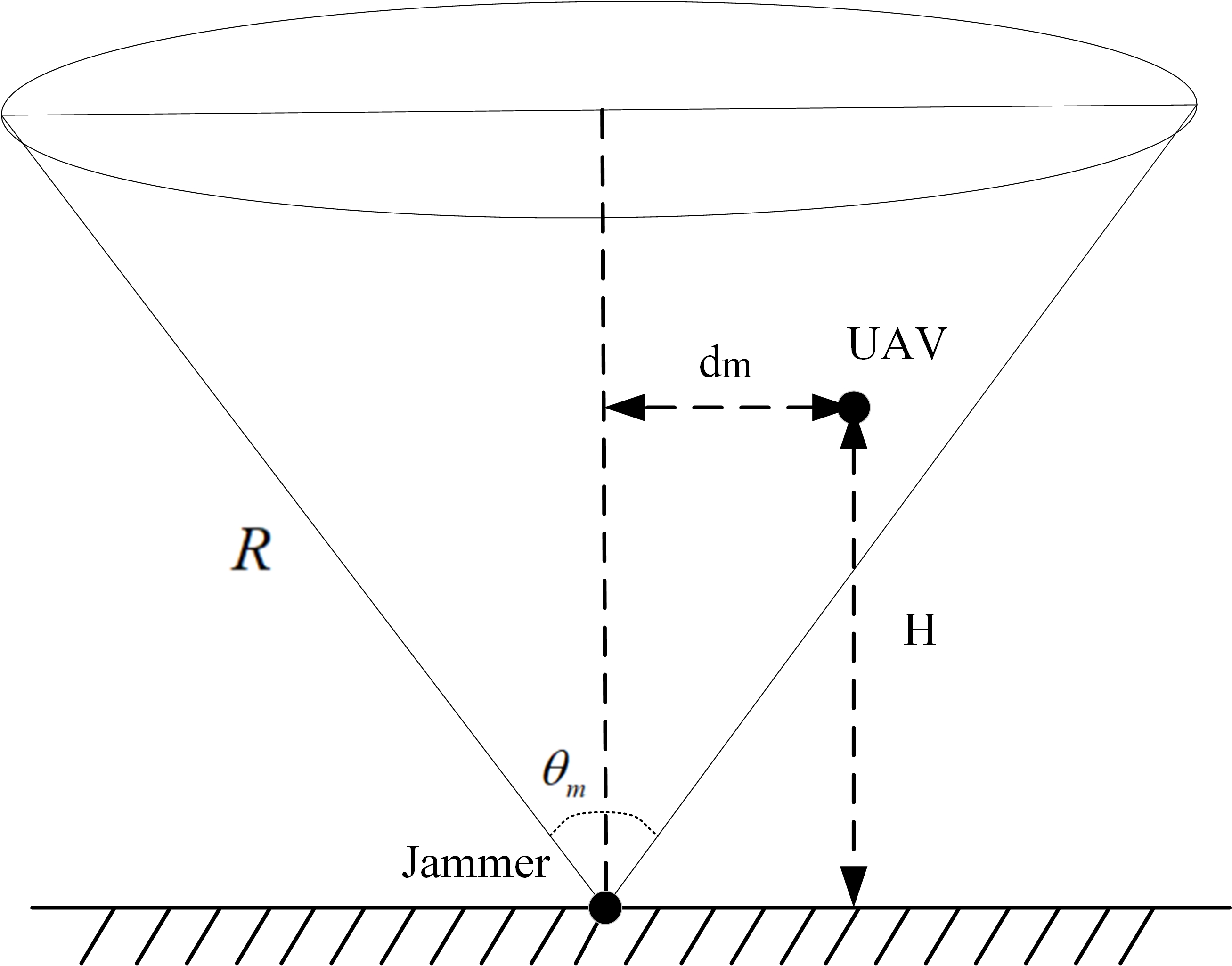}
	\caption{Interference model.}
	\label{fig3}
\end{figure}

In practical data collection scenarios, the environment may be affected by jammers. For the interference link, the path loss between the $j$-th jammer and the UAV is modeled similarly to (1) and denoted by $L_j(\delta_{n,m})$. Accordingly, the corresponding linear interference channel power gain is given by $h_j(\delta_{n,m})=10^{-L_j(\delta_{n,m})/10}$. The jammer model is designed as shown in Fig. \ref{fig3} \cite{zhang2020altitude}. It is assumed that the main lobe of each jammer’s antenna is always oriented perpendicular to the ground. The jammer’s main-lobe gain ${G}_{m}$ is predominantly confined within the beamwidth ${\theta }_{m}$, while the gain in other directions is negligible. The relationship between the isotropic antenna gain ${G}_{s}$ and the main lobe gain ${G}_{m}$ can be expressed as
\begin{equation}
	\label{eq2}
	\frac{{G}_{m}}{{G}_{s}}=\frac{\frac{P}{\pi({d}_{m}tan(\frac{{\theta }_{m}}{2}))^{2}}}{\frac{P}{4\pi{d}_{m}^{2}}}=\frac{4}{tan^{2}\frac{{\theta }_{m}}{2}},
\end{equation}
where ${d}_{m}$ represents the distance from the UAV to the centerline of the main beam. Therefore, the antenna gain ${G}_{j}({d}_{m,j})$ of the $j$-th jammer can be expressed as 
\begin{equation}
	\label{eq3}
	{G}_{j}({d}_{m,j})=\begin{cases}\frac{4{G}_{s}}{{tan}^{2}(\frac{{\theta }_{m,j}}{2})}&,{d}_{m,j}\leq Htan(\frac{{\theta }_{m,j}}{2})\\ 0&, otherwise\end{cases},
\end{equation}
where ${\theta }_{m,j}$ represents the beamwidth of the $j$-th jammer. 

Therefore, the signal-to-interference-plus-noise ratio (SINR) between the UAV and the $i$-th IoT node during the time slot ${\delta }_{n,m}$ can be expressed as
\begin{equation}
	\label{eq4}
	{SINR}_{i}({\delta }_{n,m})=\frac{{P}_{i}h_i(\delta_{n,m})}{{b}_{i}({\delta }_{n,m})S_N + \displaystyle\sum_{j=1}^{J}{P}_{j}{G}_{j}h_j(\delta_{n,m})},
\end{equation}
where ${P}_{i}$ represents the transmission power of the IoT nodes, $S_N$ represents the noise power spectral density, ${P}_{j}$ represents the transmission power of the jammer, ${b}_{i}({\delta }_{n,m})$ represents the bandwidth allocated to the $i$-th IoT node during the time slot ${\delta }_{n,m}$. The sum of the bandwidths allocated to each IoT node at any communication time slot is less than or equal to the total bandwidth size $B$, which is expressed as
\begin{equation}
	\label{eq6}
	\displaystyle\sum_{i=1}^{I}{b}_{i}({\delta }_{n,m})\leq B,\forall n\in \begin{Bmatrix}1,2,...,N\end{Bmatrix},\forall m\in \begin{Bmatrix}1,2,...,M\end{Bmatrix}.
\end{equation}

It should be noted that the SINR expression in (\ref{eq4}) represents an ideal upper-bound case, where the channel state information (CSI) is perfectly known and residual interference is not considered. This setting is adopted in the main analysis to focus on the trajectory planning and bandwidth allocation capability of the proposed framework \cite{bayerlein2020uav, wang2024trajectory}. To further evaluate the robustness of the proposed algorithm under more practical conditions, channel-gain perturbations and residual interference are additionally considered in section 4.2.8. Specifically, the desired and interference channel gains are perturbed as
\begin{equation}
	\tilde{h}_i(\delta_{n,m})=h_i(\delta_{n,m})10^{-\varepsilon_i/10},
\end{equation}
and
\begin{equation}
	\tilde{h}_j(\delta_{n,m})=h_j(\delta_{n,m})10^{\varepsilon_j/10},
\end{equation}
where $\varepsilon_i, \varepsilon_j \sim \mathcal{U}(0,\Delta_{\mathrm{CSI}})$ denote independent dB-scale channel perturbations uniformly distributed over $[0,\Delta_{\mathrm{CSI}}]$, and $\Delta_{\mathrm{CSI}}$ represents the maximum CSI uncertainty level in dB \cite{7867037}. Meanwhile, residual interference is modeled by a non-negative random coefficient $a$, satisfying $a \sim \mathrm{exp}(1/\Delta_{\mathrm{INF}})$, where $\Delta_{\mathrm{INF}}$ denotes its mean level \cite{7478654}. Accordingly, the extended SINR model used only in the additional robustness evaluation is expressed as
\begin{equation}
	SINR_i^{\mathrm{rob}}(\delta_{n,m})
	=
	\frac{P_i\tilde{h}_i(\delta_{n,m})}
	{\left[1+a\right]b_i(\delta_{n,m})S_N
		+
		\sum_{j=1}^{J}P_jG_j\tilde{h}_j(\delta_{n,m})}.
\end{equation}

The transmission rate ${C}_{i}({\delta }_{n,m})$ between the UAV and the $i$-th IoT node during the time slot ${\delta }_{n,m}$ is expressed as
\begin{equation}
	\label{eq5}
	{C}_{i}({\delta }_{n,m})={b}_{i}({\delta }_{n,m})\log_{2}{(1+{SINR}_{i}({\delta }_{n,m}))}.
\end{equation}

\subsection{UAV Energy Consumption Model}
The energy consumption of UAV mainly includes communication energy, hovering energy and propulsion energy, where the hovering energy consumption corresponds to the energy consumption when the propulsion speed is zero. In practical scenarios, the consumption of communication energy is often two orders of magnitude smaller than the propulsion energy consumption \cite{zhang2020hierarchical}. Therefore, only the UAV's hovering energy consumption and propulsion energy consumption are considered. The UAV propulsion power ${P}_{UAV}$ can be expressed as\cite{zeng2019energy}
\begin{equation}\label{eq7}
\begin{aligned}
P_{UAV}(v_u) &= P_{0}\Bigl(1 + \frac{3v_{u}^{2}}{u_{tip}^{2}}\Bigr)
  + P_{1}\Bigl(\sqrt{1 + \frac{v_{u}^{4}}{4v_{0}^{4}}}
    - \frac{v_{u}^{2}}{2v_{0}^{2}}\Bigr)^{\tfrac{1}{2}} \\
  &\quad + \frac{1}{2}d_{0}\rho s_{0}A_{r}v_{u}^{3},
\end{aligned}
\end{equation}
where ${P}_{0}$ is the blade profile power, ${v}_{u}$ is the flight speed of the UAV, ${u}_{tip}$ is the rotor tip speed of the UAV, ${P}_{1}$ is the induced power, ${v}_{0}$ is the average rotor induced speed in hover, ${d}_{0}$ is the fuselage drag ratio, $\rho$ is the air density, ${s}_{0}$ is the rotor firmness, and ${A}_{r}$ is the rotor disk area. The hovering power of the UAV is ${P}_{UAV}={P}_{0}+{P}_{1}$ when the speed of the UAV $v_u=0$.

\subsection{UAV-assisted Data Collection Optimization Model}
The volume of data transmitted during a single communication time slot $\delta_{n, m}$ is expressed as 
\begin{equation}
	\label{eq2225}
	D_{i}^{col}\left(\delta_{n,m}\right)=C_{i}\left(\delta_{n, m}\right) \delta_{n, m},
\end{equation}
Naturally, the SINR for data transmission must exceed a certain threshold. So, the expression $C_{i}\left(\delta_{n, m}\right) \delta_{n, m}$ is required to be greater than a threshold represented by $\xi$. Additionally, when the remaining data volume at $i$-th IoT node ${D}_{i}^{rem}({\delta }_{n,m})$ is less than the transmission rate ${C}_{i}({\delta }_{n,m})\delta_{n, m}$, only the remaining data volume at $i$-th IoT node can be obtained. Therefore $D_{i}^{col}({\delta }_{n,m})$ can be expressed as
\begin{equation}
	\label{eq8}
	D_{i}^{col}({\delta }_{n,m})=\begin{cases}{C}_{i}({\delta }_{n,m}){\delta }_{n,m},&  \xi \le {C}_{i}({\delta }_{n,m}){\delta }_{n,m} \\ &
    \le  {D}_{i}^{rem}({\delta }_{n,m})\\{D}_{i}^{rem}({\delta }_{n,m}),&{C}_{i}({\delta }_{n,m})\geq \\ & \max \{ \xi,{D}_{i}^{rem}({\delta }_{n,m}) \}\\
0,& otherwise\end{cases}.
\end{equation}
Our objective is to maximize the cumulative data volume collected from ground IoT nodes by optimizing the UAV's trajectory and bandwidth allocation under constraints on energy, speed, and bandwidth. The specific optimization model is formulated as follows
\begin{samepage}
\begin{equation}
P1:\max_{b_i,p_u} \quad \displaystyle\sum_{n=1}^{N}\displaystyle\sum_{m=1}^{M}\displaystyle\sum_{i=1}^{I}D_{i}^{col}({\delta }_{n,m}) \label{Problem}
\end{equation}
\begin{align}
\text {s.t.}&~E({\delta }_{n,m})\ge 0,\forall n\in \begin{Bmatrix}1,2 ... N
\end{Bmatrix},\forall m\in \begin{Bmatrix}1,2 ... M
\end{Bmatrix}  \tag{\ref{Problem}{a}} \label{Problema}\\
&~v_{u}({\delta }_{n})\in \{ 0,v \},\forall n\in \begin{Bmatrix}1,2 ... N
\end{Bmatrix}\tag{\ref{Problem}{b}} \label{Problemb}\\
&~ \displaystyle\sum_{i=1}^{I}{b}_{i}({\delta }_{n,m})\leq B,\forall n\in \begin{Bmatrix}1,2 ... N
\end{Bmatrix},\forall m\in \begin{Bmatrix}1,2 ... M
\end{Bmatrix} \tag{\ref{Problem}{c}} \label{Problemb}
\end{align}
\end{samepage}
Constraint (\ref{Problem}{a}) requires that the UAV’s energy level remains greater than 0 at all times. Constraint (\ref{Problem}{b}) requires that the UAV’s speed in each time slot must be either 0 or $v$. Constraint (\ref{Problem}{c}) requires that the total bandwidth allocated to all IoT nodes during each communication time slot must always be less than or equal to the total available bandwidth $B$.

\section{HDRL Based UAV Trajectory Design And Bandwidth Allocation}
\label{HDRL Based UAV Trajectory Design And Bandwidth Allocation}

For the optimization problem P1, each flight period contains multiple communication time slots. The action space involves two dimensions: bandwidth allocation and flight direction. We model the UAV as an intelligent agent, design its action-state space and reward function, and formulate the problem as a SMDP model. Subsequently, The TBH-DDPG algorithm is proposed to solve the model. The upper-level policy is designed for flight decision-making, while the lower-level policy makes bandwidth allocation decisions based on the flight strategy. The detailed design process of the proposed algorithm is presented in the following sections.

\subsection{Preprocessing of the State Space}
\begin{figure}[!t]
	\centering
	\includegraphics[width=3in]{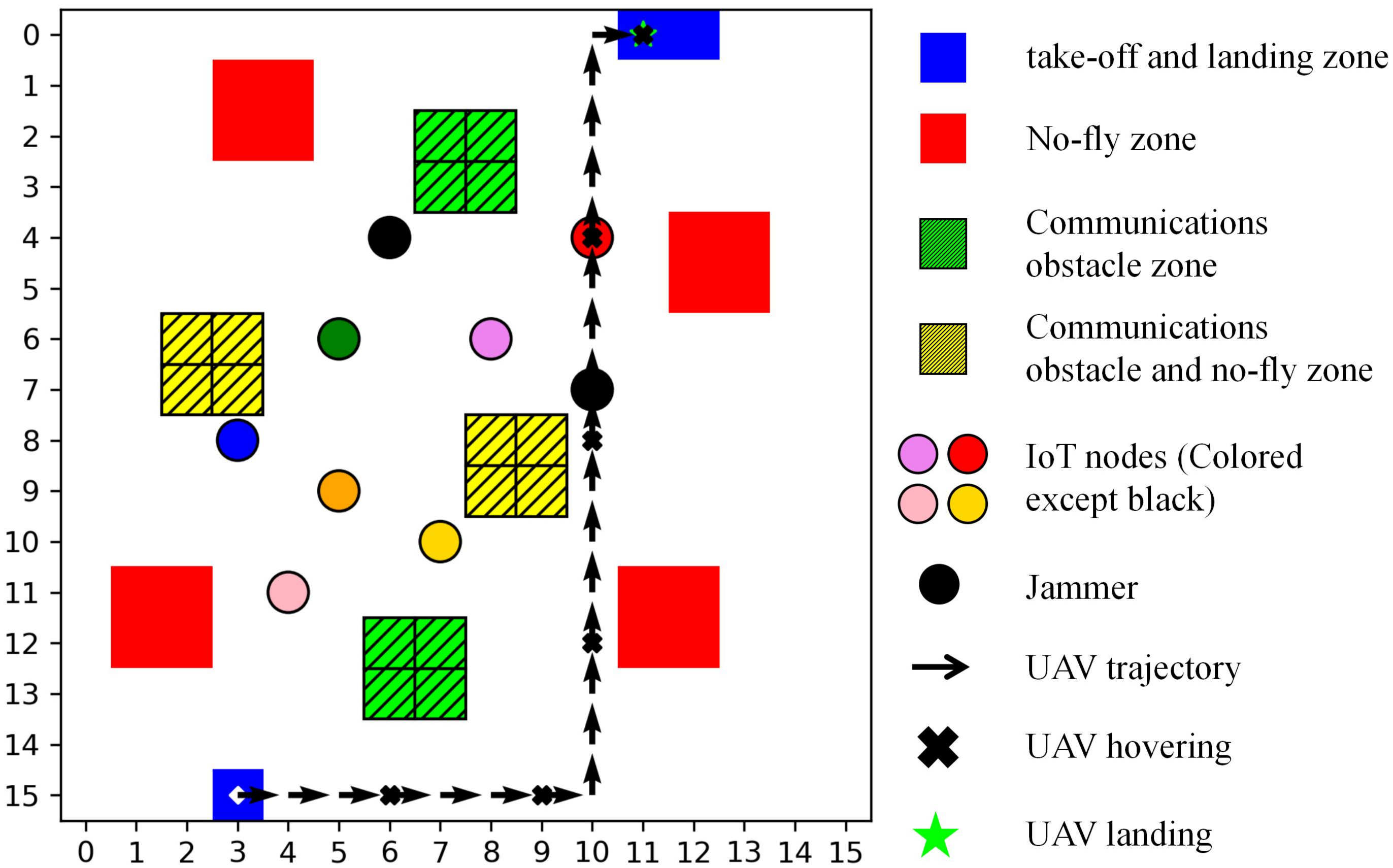}
	\caption{Scenario map of the system after abstraction.}
	\label{fig4}
\end{figure}

\begin{figure}[!t]
	\centering
	\includegraphics[width=3in]{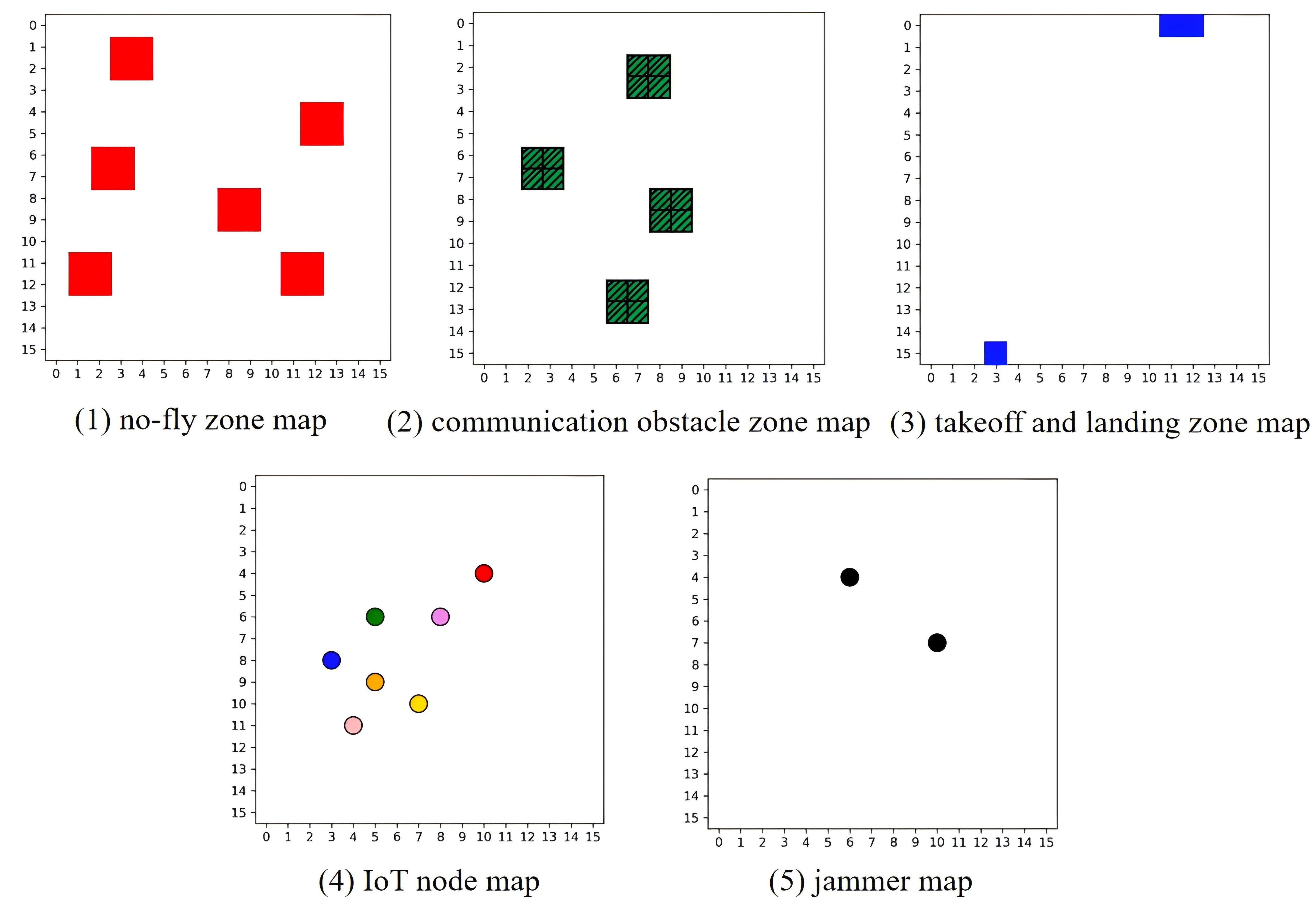}
	\caption{Five layered maps.}
	\label{fig44}
\end{figure}

The state space, as an observation of the agent, plays a critical role in the entire DRL algorithm. An abstracted map obtained by grid discretization of the specific scenario is shown in Fig. \ref{fig4}, where the considered area is divided into $Y \times Y$ cells. Here, $Y$ denotes the number of grid cells along each side and is dimensionless, while $d_{\mathrm{cell}}$ represents the side length of each grid cell in meters. In Figs. \ref{fig4} and \ref{fig44}, the value of $Y$ is set to 16, corresponding to the ranges of the x-axis and y-axis in the grid-based map. Specifically, the green communication obstacle zones force the link between the UAV and the IoT nodes to be placed in a NLoS  channel, resulting in increased signal attenuation.

The entire state space can be decomposed into five layered maps, As shown in Fig. \ref{fig44}: (1) the no-fly zone map, (2) the communication obstacle zone map, (3) the take-off and landing zone map, (4) the IoT node map, and (5) the jammer map, forming a 3D matrix represented as $\left[ Y,Y,5 \right]$. The first two dimensions of this matrix represent the position coordinates. The third dimension of the no-fly zone, communication obstacle zone, and takeoff and landing zone maps is encoded with 1 and 0 to indicate the presence or absence of a feature. Meanwhile, the IoT nodes map’s third dimension is filled with the remaining data volume of each IoT node, and the jammer map’s third dimension is filled with the calculated ratio of interference power to noise power.

The preprocessing of the state space is outlined \cite{theile2021uav}. The specific flow and propagation of parameters are shown in the state processing section of Fig. \ref{fig5}. To transform the absolute information on the map into the relative information of the UAV, the map is centralized with the UAV as the center after each acquisition of the map and UAV state information. After centralization, the map is expanded to a size of $\left[2Y-1, 2Y-1, 5\right]$ for easier feature extraction in subsequent steps. It should be noted that the expanded zones, which the UAV cannot enter, are filled with no-fly zones. For example, in Figure \ref{fig5}, the input map of size $[16*16*5]$ after applying the centering operation becomes $[31*31*5]$. The objective of the state space processing is to provide the UAV with both a general, blurred representation of the entire map and detailed information about the UAV’s immediate surroundings. This approach is consistent with the real-world observation conditions of UAVs, enabling them to focus more clearly on short-term rewards and more vaguely on long-term rewards for better performance. To achieve this, we processes the centralized map in two ways: one path performs global pooling on the map, and the other path performs central cropping. Convolutional neural networks, which are not involved in learning, are used for feature extraction from both paths. Finally, the output of the network is flattened and combined with the UAV's remaining battery information to form the input state for the algorithm.

In particular, the interference information contained in the map is also processed through this local observation mechanism. Thus, the UAV perceives the spatial interference distribution within its observation range rather than directly accessing the complete global interference environment. Meanwhile, the interference setting varies across episodes, since the transmit power and beamwidth of each jammer are randomly configured at the beginning of every episode. This design requires the UAV to make interference-aware decisions based on local observations while adapting to episode-dependent jammer characteristics.

\subsection{SMDP model}
In DRL, the MDP model is commonly used to simplify the scenario. It assumes that state transitions in the environment depend only on the previous state and is primarily composed of the components $\left \langle S, A, \mathrm{Pr}, R\right \rangle$. where $S$ represents the state space, $A$ represents the action space, $\mathrm{Pr}$ represents the probability of state transition after executing an action, and $R$ represents the reward obtained after performing an action in a given state. Since the flight time slots consist of multiple communication time slots, an extension of the MDP, SMDP, is used to model the problem. SMDP adds an option space on top of the MDP framework, which resolves the issue of inconsistent action slot contents. The specific modeling approach is as follows.

\subsubsection{State Space}
The state space is defined as $S=\{{m}_{r},{m}_{g},{m}_{b},{m}_{k},{m}_{j},{E}_{u},{p}_{u}\}$, where the first five elements represent the flight obstacle zone, communication obstacle zone, takeoff and landing zone, IoT node zone, and jammer zone mentioned in the previous section. ${E}_{u}$ represents the remaining battery level of the UAV, and ${p}_{u}$ represents the position of the UAV. The state space is obtained through the processing method described in the previous section, which results in the observation space of agent.

\begin{figure*}[!t]
	\centering
	\includegraphics[width=7in]{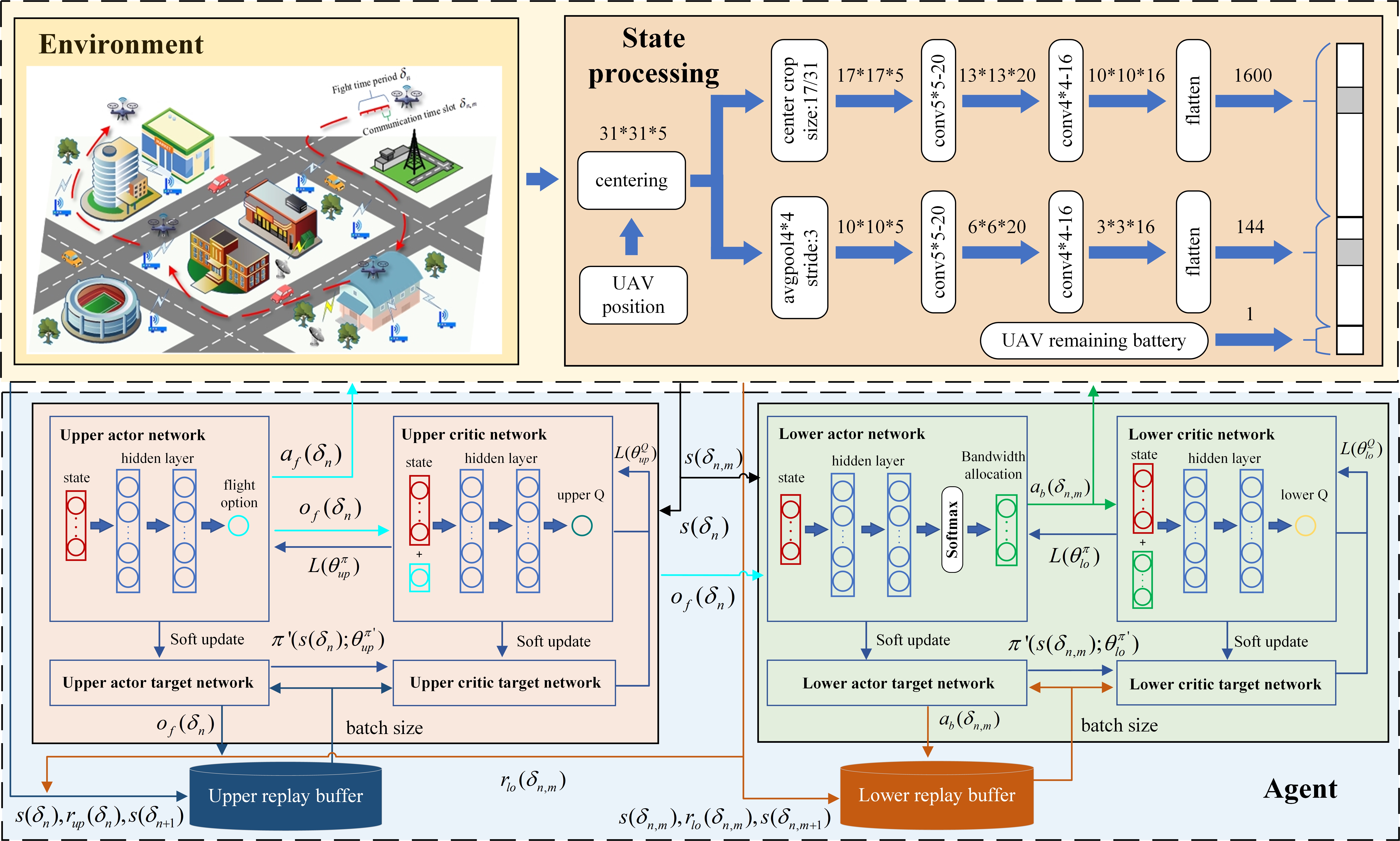}
	\caption{TBH-DDPG algorithm framework diagram.}
	\label{fig5}
\end{figure*}

\subsubsection{Action Space}
The action space is defined as $A=\{{a}_{f},{a}_{b}\}$, where ${a}_{f}$ represents the flight action space and ${a}_{b}$ represents the communication action space for bandwidth allocation. Flight action ${a}_{f}\in \begin{Bmatrix}{a}_{n},{a}_{e},{a}_{s},{a}_{w},{a}_{h},{a}_{l}\end{Bmatrix}$, where ${a}_{n}$ is one frame to the north, ${a}_{e}$ is one frame to the east, ${a}_{s}$ is one frame to the south, ${a}_{w}$ is one frame to the west, ${a}_{h}$ is hovering, and ${a}_{l}$ is landing. And the bandwidth allocation communication action ${a}_{b}=\left \langle{b}_{1},{b}_{2},\dots ,{b}_{i},\dots ,{b}_{I}\right \rangle$. That is, the total bandwidth is allocated to each node and ${b}_{i}$ is the bandwidth allocated to the $i$-th node.

\subsubsection{Option Space}
The option space is $O=\{o_f\}$, which is the space above the action. The option consists of the triad $\left \langle I,\pi,\beta \right \rangle$, where $I$ represents the initial state of the option, $\pi$ is the set of strategies under the option, and $\beta$ represents the termination condition of the option. The time slot division method is that the flight period contains multiple communication time slots, so we define ${o}_{f}\in \begin{Bmatrix}{o}_{n},{o}_{e},{o}_{s},{o}_{w},{o}_{h},{o}_{l}\end{Bmatrix}$. The initial state of the option $I$ in $o_f$ is here the state space $S$. The set of strategies $\pi =\left \langle {a}_{f},{a}_{b1},{a}_{b2},\dots ,{a}_{bm},\dots ,{a}_{bM}\right \rangle$, where ${a}_{f}$ corresponds to the selected ${o}_{f}$. For example, when the upper-level policy selects ${o}_{n}$, the policy set becomes $\pi =\left \langle {a}_{n},{a}_{b1},{a}_{b2},\dots ,{a}_{bm},\dots ,{a}_{bM}\right \rangle$, meaning that by selecting ${o}_{f}$, the choice of ${a}_{f}$ is also completed. And ${a}_{bm}$ refers to the bandwidth allocation communication operation carried out in the $m$-th communication time slot. Finally, the termination condition $\beta$ is that all actions in $\pi$ are completed sequentially.

\subsubsection{Reward Design}
The reward consists of two parts. First, the lower-level reward obtained during the communication time slot is
\begin{equation}
	\label{eq10}
	{r}_{lo}({\delta }_{n,m})={r}_{collection}({\delta }_{n,m})+{r}_{loss}({\delta }_{n,m}),
\end{equation}
where ${r}_{collection}({\delta }_{n,m})$ represents the data collection reward, and ${r}_{loss}({\delta }_{n,m})$ represents the data loss penalty. The data collection reward is achieved after each data collection operation, calculated as the amount of data collected multiplied by a fixed proportion. Specifically, it is defined as
\begin{equation}
	\label{eq11}
	{r}_{collection}({\delta }_{n,m})={\epsilon }_{cen} \cdot {r}_{cen}({\delta }_{n,m}),
\end{equation}
where ${\epsilon }_{cen}$ is the proportional coefficient for the data collection reward, and ${r}_{cen}({\delta }_{n,m})=\displaystyle\sum_{i=1}^{I}({D}_{i}({\delta }_{n,m})-{D}_{i}({\delta }_{n,m+1}))$ is the amount of data collected during a single communication time slot. The data loss penalty is applied when the data storage exceeds the storage capacity limit of the IoT nodes, and a fixed penalty value is assigned. That is
\begin{equation}
	\label{eq12}
	{r}_{loss}({\delta }_{n,m})=\begin{cases} {{r}_{ls}}&,if\ {D}_{i}({\delta }_{n,m})\ge {D}_{max}\\ 0&, otherwise\end{cases},
\end{equation}
where ${r}_{ls}$ represents the value of the penalty constant for data loss.

There is also the upper-level reward obtained during the flight time period. It includes the rewards from all communication time slots within the flight time slot and the reward obtained from executing the flight action. That is
\begin{equation}
	\label{eq13}
	{r}_{up}({\delta }_{n})={r}_{f}({\delta }_{n})+{r}_{lo}({\delta }_{n,1})+...+{r}_{lo}({\delta }_{n,m})+...+{r}_{lo}({\delta }_{n,M}),
\end{equation}
where ${r}_{f}({\delta }_{n})$ represents the sum of the collision penalty, the return penalty, and the crash penalty. That is
\begin{equation}
	\label{eq131}
	{r}_{f}({\delta }_{n})={r}_{collision}({\delta }_{n})+{r}_{return}({\delta }_{n})+{r}_{nland}({\delta }_{n}).
\end{equation}
The collision penalty is applied when the UAV enters a no-fly zone, assigning a fixed penalty value. The specific expression is as follows.
\begin{equation}
	\label{eq14}
	{r}_{collision}({\delta }_{n})=\begin{cases}{r}_{csn}&,if\ {p}_{u}({\delta }_{n})\ in\ red\ zone\\ 0&, otherwise\end{cases},
\end{equation}
where ${r}_{csn}$ represents the collision penalty constant. The return penalty is applied when the UAV's battery level falls below a threshold. The penalty increases as the battery level decreases and the distance from the landing point increases. That is
\begin{equation}
	\label{eq15}
	{r}_{return}({\delta }_{n})=\begin{cases}{d}_{re}({\delta }_{n})\cdot ({\epsilon }_{re}-E({\delta }_{n}))&,if\ E({\delta }_{n})\le {E}_{tsd}\\ 0&, otherwise\end{cases},
\end{equation}
where ${d}_{re}({\delta }_{n})$ is the current distance of the UAV from the take-off and landing zone, ${\epsilon }_{re}$ is the return penalty coefficient, $E({\delta }_{n})$ is the UAV's current remaining battery level, and ${E}_{tsd}$ is the return penalty threshold. Finally, the landing penalty is applied when the UAV's battery is depleted and it has not yet landed. That is
\begin{equation}
	\label{eq16}
	{r}_{nland}({\delta }_{n})=\begin{cases}{\epsilon }_{ld1} \cdot {d}_{re}({\delta }_{n})+{\epsilon }_{ld2}&,if\ E({\delta }_{N,M})\le 0\ \\ & and\ not\ landed\\ 0&, otherwise\end{cases},
\end{equation}
where ${\epsilon }_{ld1}$ and ${\epsilon }_{ld2}$ are the penalty coefficients for not landed.

\subsubsection{Process Design for a Single Option}
The execution process of a single option is described in the following. First, an option $o_f({\delta }_{n})$ is output by the upper-level network based on the observed state $s({\delta }_{n})$. The agent then executes the flight action $a_f({\delta }_{n})$ according to this option $o_f({\delta }_{n})$, achieving the flight reward $r_f({\delta }_{n})$. Next, the process enters bandwidth allocation slots. The bandwidth allocation action $a_b({\delta }_{n,1})$ is output by the lower-level network based on the observed state $s({\delta }_{n,1})$. The agent then executes the bandwidth allocation action $a_b({\delta }_{n,1})$ to collect data volume from the IoT nodes, achieving the lower‐level network’s reward $r_{lo}({\delta }_{n,1})$, as expressed in (\ref{eq10}). This bandwidth‐allocation decision and execution cycle repeats until the current option terminates. Finally, the sum of the flight reward and all accumulated lower‐level rewards constitutes the upper‐level network’s reward $r_{up}({\delta }_{n})$, as expressed in (\ref{eq13}).

\subsection{Design of the TBH-DDPG algorithm}
The TBH-DDPG algorithm is designed to solve the SMDP model, where the upper-level networks are used for trajectory optimization and the lower-level networks are used for bandwidth allocation optimization. Moreover, the upper-level flight action is a discrete action, while the lower-level bandwidth allocation action is a continuous action. The details are presented below.

\subsubsection{DDPG Algorithm}
A brief introduction to the deep deterministic policy gradients (DDPG) algorithm \cite{10475324} framework is provided, focusing primarily on the upper‐level neworks as an illustrative example. The DDPG algorithm consists mainly of the actor and critic networks. The actor network formulates a policy based on the input state $s({\delta }_{n})$, represented as ${\pi }_{up}(s({\delta }_{n});{\theta }_{up}^{\pi })$, where ${\theta }_{up}^{\pi }$ is the neural network parameters, while the critic network evaluates the value of the action ${a}_{f}({\delta }_{n})$ taken in state $s({\delta }_{n})$, represented as ${Q_{up}}(s({\delta }_{n}),{a}_{f}({\delta }_{n});{\theta }^{Q})$. Since the actor network can directly output the action value, DDPG can handle both continuous and discrete action problems. 

DDPG uses replay buffer and target network techniques in order to stabilize the learning process. The replay buffer stores the state, action, reward, and next state of each step throughout the training process, and then randomly samples a batch of size ${S}_{batch}$ from it each time to train the network. Stabilizing the training process by increasing sample reuse and decreasing correlation between samples. The target network technique is to use the same network architecture ${\pi }_{up}(s({\delta }_{n});{\theta }_{up}^{\pi '})$ and ${Q_{up}}(s({\delta }_{n}),{a}_{f}({\delta }_{n});{\theta }_{up}^{Q'})$ as ${\pi }_{up}(s({\delta }_{n});{\theta }_{up}^{\pi })$ and ${Q_{up}}(s({\delta }_{n}),{a}_{f}({\delta }_{n});{\theta }_{up}^{Q})$, and to prevent training instability due to large changes in the network parameters by means of soft updates. The expression for the soft update is
\begin{equation}
	\label{eq17}
	{\theta }^{'}=\varepsilon \theta + (1 - \varepsilon ){\theta }^{'},
\end{equation}
where ${\theta }^{'}$ represents the target neural network parameters, and $\theta$ represents the neural network parameters.

Next, the update methods for the two networks are described, namely their loss functions. For the actor network, it aims to maximize the Q-value of the chosen policy at state $s({\delta }_{n})$, Therefore, its loss function is defined as follows.
\begin{equation}
	\label{eq18}
	L({\theta }_{up}^{\pi })={\mathbb{E}}_{{S}_{batch}}[-Q_{up}(s({\delta }_{n}),\pi_{up} (s({\delta }_{n});{\theta }_{up}^{\pi });{\theta }_{up}^{Q})],
\end{equation}
where ${\mathbb{E}}_{{S}_{batch}}[\cdot ]$ represents taking the mean of the values obtained for the selected batch. For the critic network, it aims to evaluate the accumulated reward that can be obtained after executing action ${a}_{f}({\delta }_{n})$ at state $s({\delta }_{n})$, and strives to make this value as close as possible to the true value. Its loss function is defined as follows.
\begin{equation}
	\label{eq19}
	L({\theta }_{up}^{Q })={\mathbb{E}}_{{S}_{batch}}[{(Q_{up}(s({\delta }_{n}),{a}_{f}({\delta }_{n});{\theta }_{up}^{Q })-T{a}_{f}({\delta }_{n}))}^{2}],
\end{equation}
where 
\begin{equation}
	\label{eq190}
        T{a}_{f}({\delta }_{n})={r}_{up}({\delta }_{n})+\gamma {Q}_{up}(s({\delta }_{n+1}),\pi (s({\delta }_{n+1});{\theta }_{up}^{\pi '});{\theta }_{up}^{Q'}))
\end{equation}
represents the update target.

\subsubsection{TBH-DDPG Algorithm}
The TBH-DDPG algorithm is designed by integrating DDPG, with the overall framework shown in Fig. \ref{fig5} and Algorithm 1. First, the lower-level actor network is designed to output the proportion coefficient for the bandwidth allocation of each IoT node. Therefore, the lower-level actor network first outputs $I$ values, which are then forwarded through a softmax layer to ensure that their sum equals 1. The result is represented by $\mathbf{{\lambda}_{b}}=\left \langle {\lambda}_{b1},{\lambda}_{b2},\dots ,{\lambda}_{bi},\dots ,{\lambda}_{bI}\right \rangle$. The bandwidth allocation size for each IoT node is represented as ${a}_{b}=B \cdot \mathbf{{\lambda}_{b}}$.

In the entire algorithm process, the first step is to initialize the network parameters, the environment map, the UAV's position, and the UAV battery level. Then, the flight period begins. The environment map is propagated forward to obtain the input state $s({\delta }_{n})$. The upper-level network obtains the option ${o}_{f}({\delta }_{n})$ and flight action ${a}_{f}({\delta }_{n})$ according to the input state $s({\delta }_{n})$. Then, the UAV directly executes the action, achieving the reward ${r}_{up}({\delta }_{n})$ for the flight action and updating environment map.

Next, the algorithm transitions into the communication time slot. The environment map is propagated forward to obtain state $s({\delta }_{n,m})$. The lower-level network then uses $s({\delta }_{n,m})$ to obtain the bandwidth allocation action $a_b({\delta }_{n,m})$. The UAV performs the bandwidth allocation and subsequently communicates with the IoT nodes, thereby achieving the reward $r_{lo}({\delta }_{n,m})$ and updating environment map. The next environment map is then propagated forward to obtain $s({\delta }_{n,m+1})$. This process continues in a loop until the communication time slots are completed.

Finally, during the entire process, the upper-level replay buffer stores $\left \langle s({\delta }_{n}),{a}_{f}({\delta }_{n}),{r}_{up}({\delta }_{n}),s({\delta }_{n+1})\right \rangle$, and the lover-level replay buffer stores $\left \langle s({\delta }_{n,m}),{a}_{b}({\delta }_{n,m}),{r}_{lo}({\delta }_{n,m}),s({\delta }_{n,m+1})\right \rangle$. The network is updated according to (\ref{eq17}), (\ref{eq18}) and (\ref{eq19}). The proposed algorithm ultimately enables the UAV to maximize the cumulative volume of data collected from distributed IoT nodes.

\begin{algorithm}[htbp]
\caption{TBH-DDPG}\label{TBH-DDPG}
\begin{algorithmic}[1]
\STATE Initialize all networks, including the upper actor network ${\pi }_{up}(s;{\theta }_{up}^{\pi })$ and its target actor network${\pi }_{up}(s;{\theta }_{up}^{\pi'})$, the upper critic network ${Q}_{up}(s;{\theta }_{up}^{Q })$ and its target critic network ${Q}_{up}(s;{\theta }_{up}^{Q'})$, the lower actor network ${\pi }_{lo}(s;{\theta }_{lo}^{\pi })$ and its target network ${\pi }_{lo}(s;{\theta }_{lo}^{\pi '})$, the lower critic network ${Q}_{lo}(s,a;{\theta }_{lo}^{Q })$ and its target network ${Q}_{lo}(s,a;{\theta }_{lo}^{Q'})$
\STATE Initialize the upper replay buffer of size U and the lower replay buffer of size L
\FOR{each episode}
    \STATE Initialize the environment maps, UAV's position and electricity ${E}_{max}$
    \FOR{each fly time slot $n$}
        \STATE the environment maps forward propagation to get the input state $s({\delta }_{n})$
        \STATE Input the state $s({\delta }_{n})$ into the upper actor network to get ${o}_{f}({\delta }_{n})$ and ${a}_{f}({\delta }_{n})$
        \STATE UAV executes action ${a}_{f}({\delta }_{n})$ and achives a flight reward ${r}_{up}({\delta }_{n})$, and the next environment map
        \FOR{each communication time slot $m$}
            \STATE Environment maps are forward propagated to get $s({\delta }_{n,m})$
            \STATE Input $s({\delta }_{n,m})$ into the lower actor network and add noise to get the bandwidth allocation communication action ${a}_{b}({\delta }_{n,m})$
            \STATE UAV executes action ${a}_{b}({\delta }_{n,m})$ and achives reward ${r}_{lo}({\delta }_{n,m})$
            \STATE IoT node data increased, ${D}_{i}({\delta }_{n,m})={D}_{i}({\delta }_{n,m})+{d}_{inc}$. Then update environment maps
            \STATE Adding reward to the upper network, ${r}_{up}({\delta }_{n})={r}_{up}({\delta }_{n})+{r}_{lo}({\delta }_{n,m})$
            \STATE Environment maps are forward propagated to get $s({\delta }_{n,m+1})$
            \STATE Store $\left \langle s({\delta }_{n,m}),{a}_{b}({\delta }_{n,m}),{r}_{lo}({\delta }_{n,m}),s({\delta }_{n,m+1})\right \rangle$ in the lower replay buffer
        \ENDFOR
        \STATE Environment maps are forward propagated to get $s({\delta }_{n+1})$
        \STATE Store $\left \langle s({\delta }_{n}),{a}_{f}({\delta }_{n}),{r}_{up}({\delta }_{n}),s({\delta }_{n+1})\right \rangle$ in the upper replay buffer
        \IF{up to upper replay buffer storage limit U}
            \STATE Update the upper network using (\ref{eq18}), (\ref{eq19})
            \STATE Update the upper target network using (\ref{eq17})
        \ENDIF
        \IF{up to lower replay buffer storage limit L}
            \STATE Update the lower network using (\ref{eq18}), (\ref{eq19})
            \STATE Update the lower target network using (\ref{eq17})
        \ENDIF
    \ENDFOR
\ENDFOR
\end{algorithmic}
\end{algorithm}

\section{PERFORMANCE EVALUATION}
In this section, the performance of the proposed TBH-DDPG algorithm is evaluated. The TBH-DDPG algorithm is implemented based on the PyTorch framework with torch version 1.9.1, and the graphics card used is an RTX 4060Ti.

\subsection{Environmental parameters}
The environment parameters are listed in Table \ref{tab2}. The parameters for IoT nodes, zones, and jammers in different scenarios are provided in Table \ref{tab3}. The primary differences among the scenarios lie in the locations of the UAV’s take-off and landing zones,  IoT nodes and various obstacle zones. The whole map is shown in Fig. \ref{fig4}, where each grid cell on the map has a width and length of 20 meters. Additionally, in a single episode, the jammer power and beamwidth are randomly selected from $\left \{10,20,30,40,50 \right \}$ and $\left \{30,60,90\right \}$, respectively. Once selected, these values remain unchanged throughout the episode. Based on Table \ref{tab2}, the calculated propulsion energy consumption is 178.2958w and the hovering energy consumption is 168.4642w. For ease of implementation in the program, these values are normalized: the hovering energy consumption is set to 1, and the propulsion energy consumption is set to 1.0582.

\begin{table}[htbp]
\caption{Environmentally Relevant Parameters}\label{tab2}
\centering
\renewcommand{\arraystretch}{1.2} 
\begin{tabular}{|c|c|c|}
\hline
\textbf{Variables} & \textbf{Explanation}                             & \textbf{Value} \\ \hline
$Y$ & Number of grid cells per side & 16 \\ \hline
$d_{\mathrm{cell}}$ & Grid cell side length (m) & 20 \\ \hline
${P}_{i}$   & Transmit power of the IoT node(w)      & 0.1          \\ \hline
${d}_{inc}$ & IoT node data growth volume(Mb)        & 0.2          \\ \hline
${D}_{max}$ & IoT node data storage limit(Mb)        & 65           \\ \hline
${S}_{N}$   & Noise power spectral density(w/Hz) & 1e-17        \\ \hline
$\xi$       & Communication rate threshold       & 0.001        \\ \hline
$M$ & Number of communications                   & 4            \\ \hline
$B$ & Total allocable bandwidth size(MHz)        & 0.5          \\ \hline
${\alpha }_{LoS}$       & LoS fading factor      & 2.27    \cite{bayerlein2020uav}     \\ \hline
${\alpha }_{NLoS}$      & NLoS fading factor     & 3.64   \cite{bayerlein2020uav}      \\ \hline
${\sigma }_{LoS}^{2}$   & variance of the shadowing effect of LoS  & 2 \cite{bayerlein2020uav}\\ \hline
${\sigma }_{NLoS}^{2}$  & variance of the shadowing effect of NLoS & 5 \cite{bayerlein2020uav}\\ \hline
$H$         & UAV flight altitude(m)           & 30             \\ \hline
${E}_{max}$ & UAV Normalized Power             & 90             \\ \hline
$v$         & UAV flying speed(m/s)            & 20             \\ \hline
${P}_{0}$   & blade profile power(w)           & 79.85          \\ \hline
${P}_{1}$   & induced power(w)                 & 88.62          \\ \hline
${u}_{tip}$ & rotor tip speed of the UAV(m/s)  & 120            \cite{zhang2020hierarchical} \\ \hline
${v}_{0}$   & average rotor induced speed in hover(m/s)  & 4.03 \cite{zhang2020hierarchical}\\ \hline
${d}_{0}$   & fuselage drag ratio              & 0.6            \cite{zhang2020hierarchical}\\ \hline
$\rho $     & the air density(kg/m)            & 1.225          \cite{zhang2020hierarchical}\\ \hline
${s}_{0}$   & rotor firmness                   & 0.05           \cite{zhang2020hierarchical}\\ \hline
${A}_{r}$   & rotor disk area(m$^2$)           & 0.503          \cite{zhang2020hierarchical}\\ \hline
\end{tabular}
\end{table}

\begin{table*}[!t]
\caption{Scenario Parameters}\label{tab3}
\centering
\renewcommand{\arraystretch}{1.2} 
\begin{tabular}{|c|c|c|c|}
\hline
           & IoT nodes coordinates                                   & IoT nodes initial data volume (Mb)  & Jammers coordinates \\ \hline
Scenario 1 & \{(4,11),(3,8),(5,9),(5,6),(7,10),(10,4),(8,6)\}  & {[}50,50,10,10,50,50,10{]} & \{(6,4),(10,7)\}  \\ \hline
Scenario 2 & \{(3,10),(6,12),(8,10),(7,7),(10,9),(9,3),(6,5)\} & {[}50,50,10,10,50,50,10{]} & \{(6,9),(10,7)\}  \\ \hline
Scenario 3 & \{(9,13),(8,10),(6,3),(5,6),(5,10),(9,4),(3,2)\}  & {[}10,10,10,50,50,50,50{]} & \{(4,4),(7,12)\}  \\ \hline
\end{tabular}
\end{table*}

The relevant parameter settings for the algorithm's network are listed in Table \ref{tab4}.

\begin{table}[htbp]
\caption{Network Parameters}\label{tab4}
\centering
\renewcommand{\arraystretch}{1.2} 
\begin{tabular}{|c|c|}
\hline
\textbf{Algorithmic parameters}             & \textbf{Value}        \\ \hline
batchsize                                   & 128                   \\ \hline
hidden layer size                           & 128                   \\ \hline
upper replay buffer size                    & 30000                 \\ \hline
upper discount factor $\gamma$              & 0.99                  \\ \hline
upper soft update parameter $\epsilon$      & 0.005                 \\ \hline
upper actor learning rate                   & 0.0001                \\ \hline
upper critic learning rate                  & 0.0001                 \\ \hline
upper exploration noise                  & 3                 \\ \hline
lower replay buffer size                    & 30000                 \\ \hline
lower discount factor $\gamma$              & 0.995                 \\ \hline
lower soft update parameter $\epsilon$      & 0.00001               \\ \hline
lower actor learning rate                   & 0.0001                \\ \hline
lower critic learning rate                  & 0.0001                \\ \hline
lower exploration noise                  & 1                 \\ \hline
data collection reward factor ${\epsilon }_{cen}$             & 1.5                 \\ \hline
data loss penalty ${r}_{ls}$                                  & -1                  \\ \hline
collision penalty ${r}_{csn}$                                 & -7                  \\ \hline
return penalty factor ${\epsilon }_{re}$                      & 10                  \\ \hline
return electricity threshold ${E}_{tsd}$                      & 10                  \\ \hline
landing penalty factor 1 ${\epsilon }_{ld1}$                  & -10                 \\ \hline
landing penalty factor 2 ${\epsilon }_{ld2}$                  & -100                \\ \hline
\end{tabular}
\end{table}

\subsection{Performance and Analysis}
The main algorithms compared include a trajectory and bandwidth joint network optimization algorithm based on DDPG (TBJN-DDPG), a trajectory and bandwidth allocation optimization algorithm based on hierarchical soft actor-critic (TBH-SAC) and a time-division multiple access-based UAV data collection algorithm using DDPG (TDMA-DDPG):

\textbf{TBJN-DDPG} \cite{ding20203d}: A non-hierarchical trajectory optimization and bandwidth allocation algorithm, which employs a single-layer joint action network to simultaneously output both flight and bandwidth allocation actions.

\textbf{TBH-SAC} \cite{fan2022ris}: A framework similar to the proposed algorithm, but replacing the base algorithm DDPG with the SAC algorithm. Specifically, the lower-level actor network outputs the mean and variance of the bandwidth allocation actions for all IoT nodes, forming a normal distribution. Bandwidth allocation ratios are then obtained by sampling from this distribution and applying the softmax function.

\textbf{TDMA-DDPG} \cite{bayerlein2020uav}: A path planning algorithm based on DDPG. For bandwidth allocation, it adopts a TDMA approach by assigning all available bandwidth to the IoT nodes with the highest calculated communication rate at the current moment, thereby maximizing data collection.

\subsubsection{\textbf{Reward Convergence Performance}}
\begin{figure}[!t]
	\centering
	\includegraphics[width=3.3in]{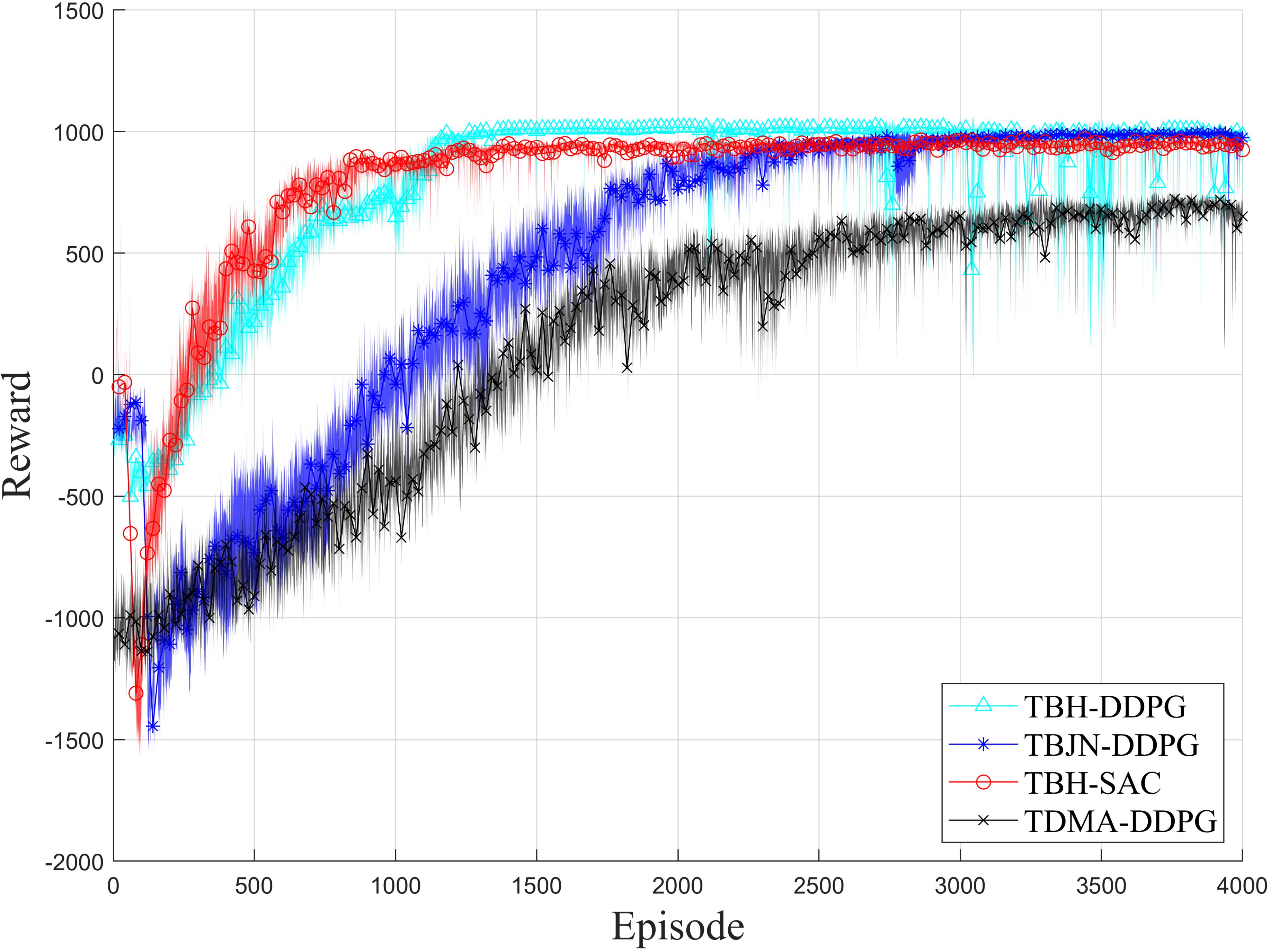}
	\caption{Reward training curves.}
	\label{fig6}
\end{figure}

As shown in Fig. \ref{fig6}, each algorithm was run five times. The solid lines represent the average rewards of the five experiments, while the shaded areas indicate the range of the mean plus and minus $50\%$ of the variance. From Fig. \ref{fig6}, the proposed TBH-DDPG algorithm converges approximately within 1100–1400 episodes, while the non-hierarchical TBJN-DDPG algorithm converges approximately within 2100–2400 episodes. To quantify the convergence-speed improvement, the midpoint of each convergence interval is used for comparison. Specifically, the midpoint convergence episode of TBH-DDPG is 1250, while that of TBJN-DDPG is 2250. Therefore, the convergence-speed improvement of the proposed algorithm over TBJN-DDPG is calculated as $44.44\%$. This demonstrates that the proposed algorithm attains an optimal convergence value while ensuring a rapid convergence rate. The hierarchical algorithm TBH-SAC demonstrates slightly lower performance than the proposed algorithm and TBJN-DDPG. But it achieves a comparable convergence speed to the proposed algorithm. 

The ultimate goal of reinforcement learning is to maximize rewards, making efficient reward design crucial for the algorithm. In the proposed hierarchical algorithm, the lower-level rewards include only data collection and data loss rewards, and the optimization is limited to bandwidth allocation actions. The upper-level rewards include the lower-level rewards, but optimize only the flight options. In comparison, the non-hierarchical algorithm considers all rewards and simultaneously optimizes both flight and bandwidth allocation actions. Therefore, the proposed algorithm effectively reduces convergence time compared to the non-hierarchical approach. Moreover, the algorithm endowed with autonomous bandwidth allocation consistently outperforms the TDMA-DDPG approach in terms of converged reward. Through reasonable bandwidth allocation, the algorithm can collect data volume from multiple IoT nodes, significantly improving bandwidth utilization efficiency.

\subsubsection{\textbf{UAV Trajectories, Cumulative Data Collection, and Bandwidth Allocation Ratios}} 
\begin{figure*}[!t]
    \centering
    \subfloat[]{
    		\includegraphics[width=1.7in]{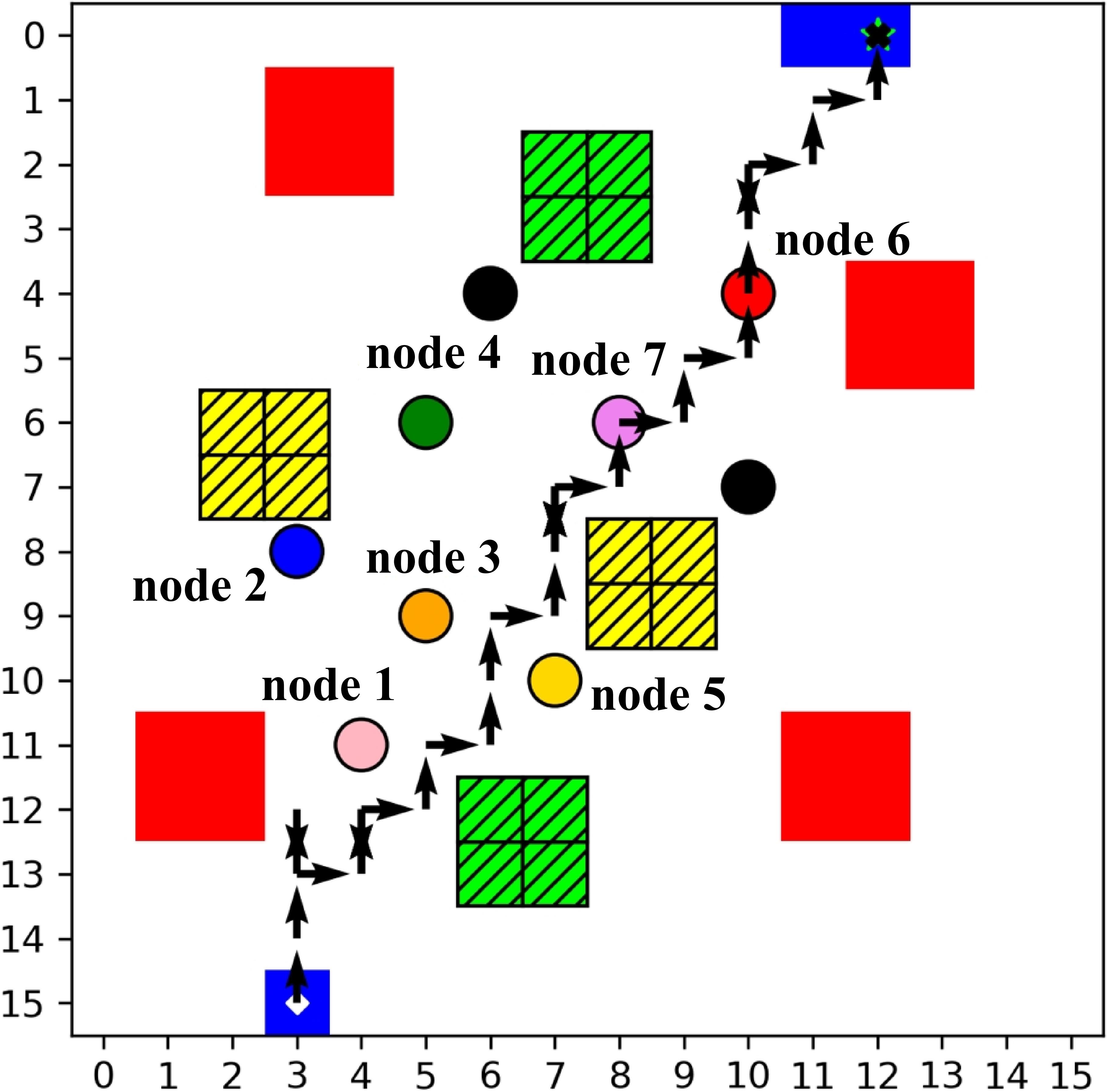}}
    \subfloat[]{
    		\includegraphics[width=2.2in]{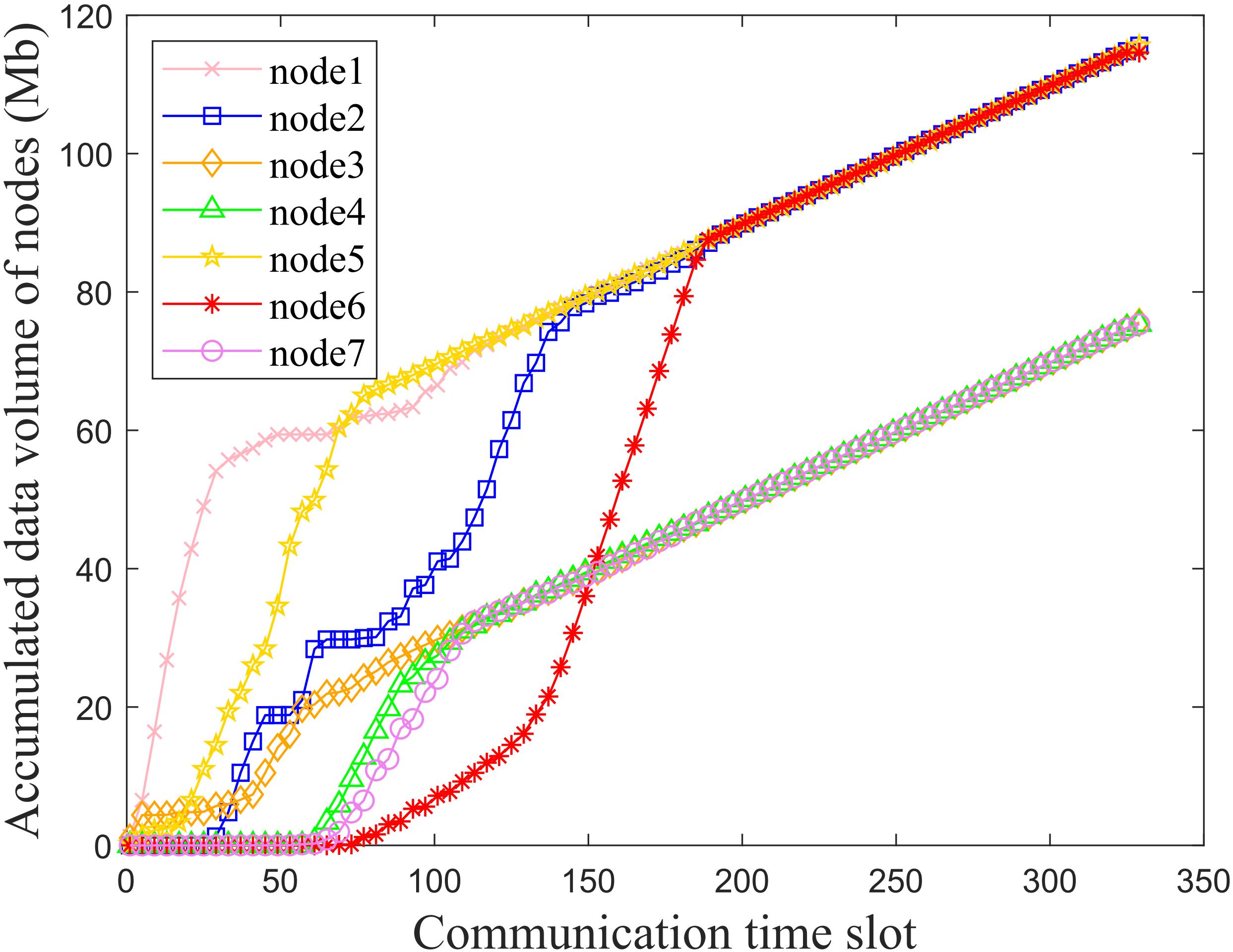}}
    \subfloat[]{
    		\includegraphics[width=2.4in]{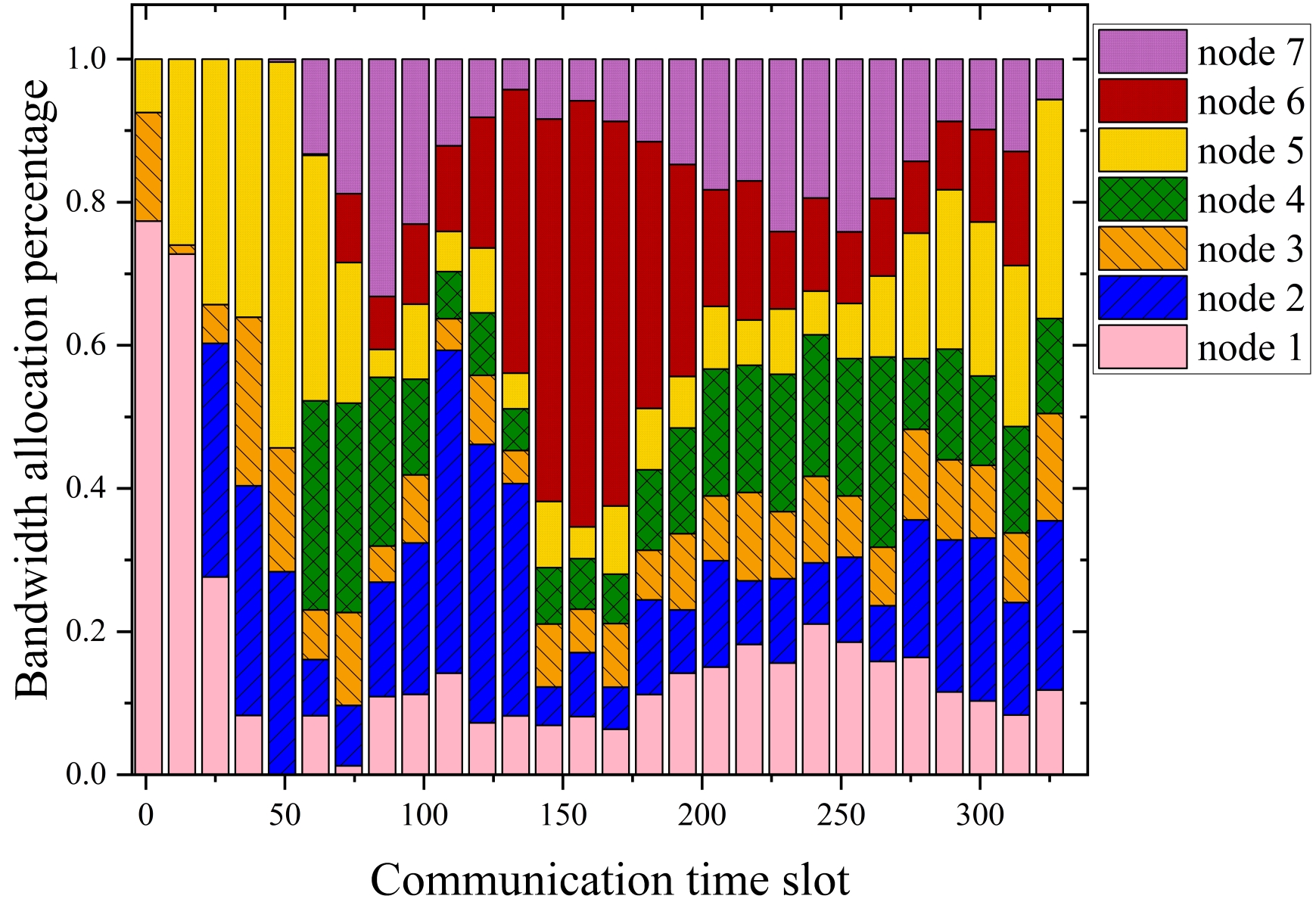}}
    \qquad
    \centering
    \subfloat[]{
    		\includegraphics[width=1.7in]{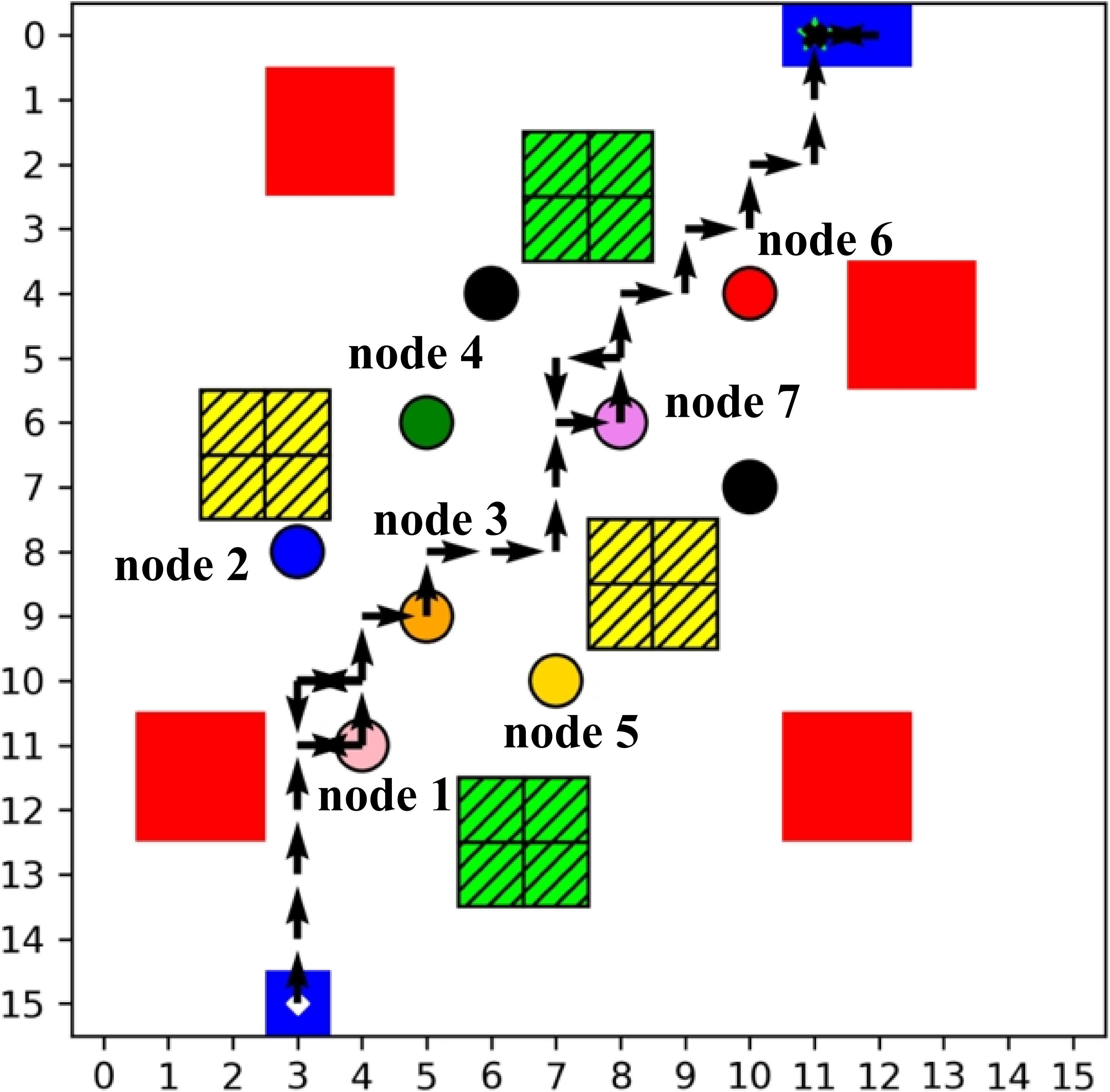}}
    \subfloat[]{
    		\includegraphics[width=2.2in]{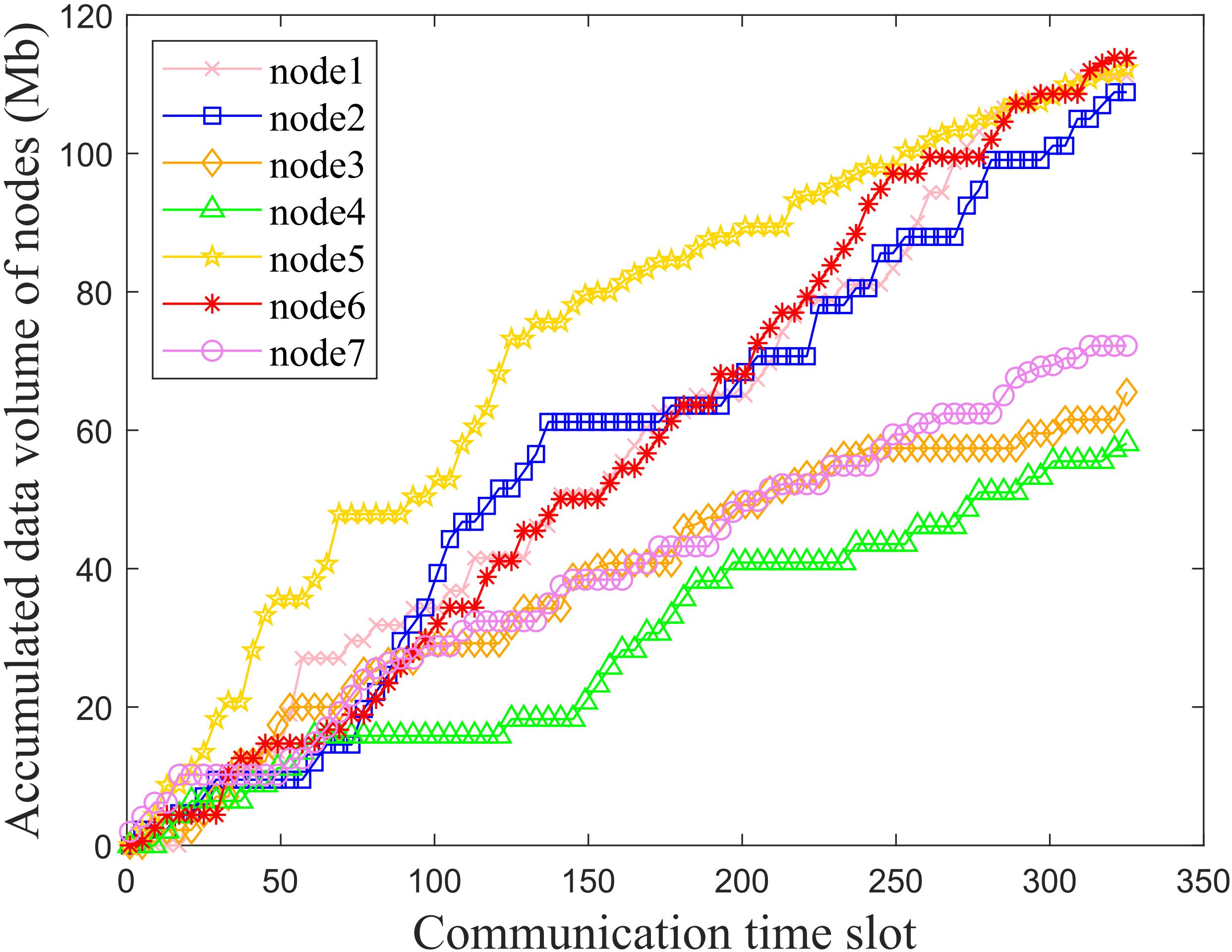}}
    \subfloat[]{
    		\includegraphics[width=2.4in]{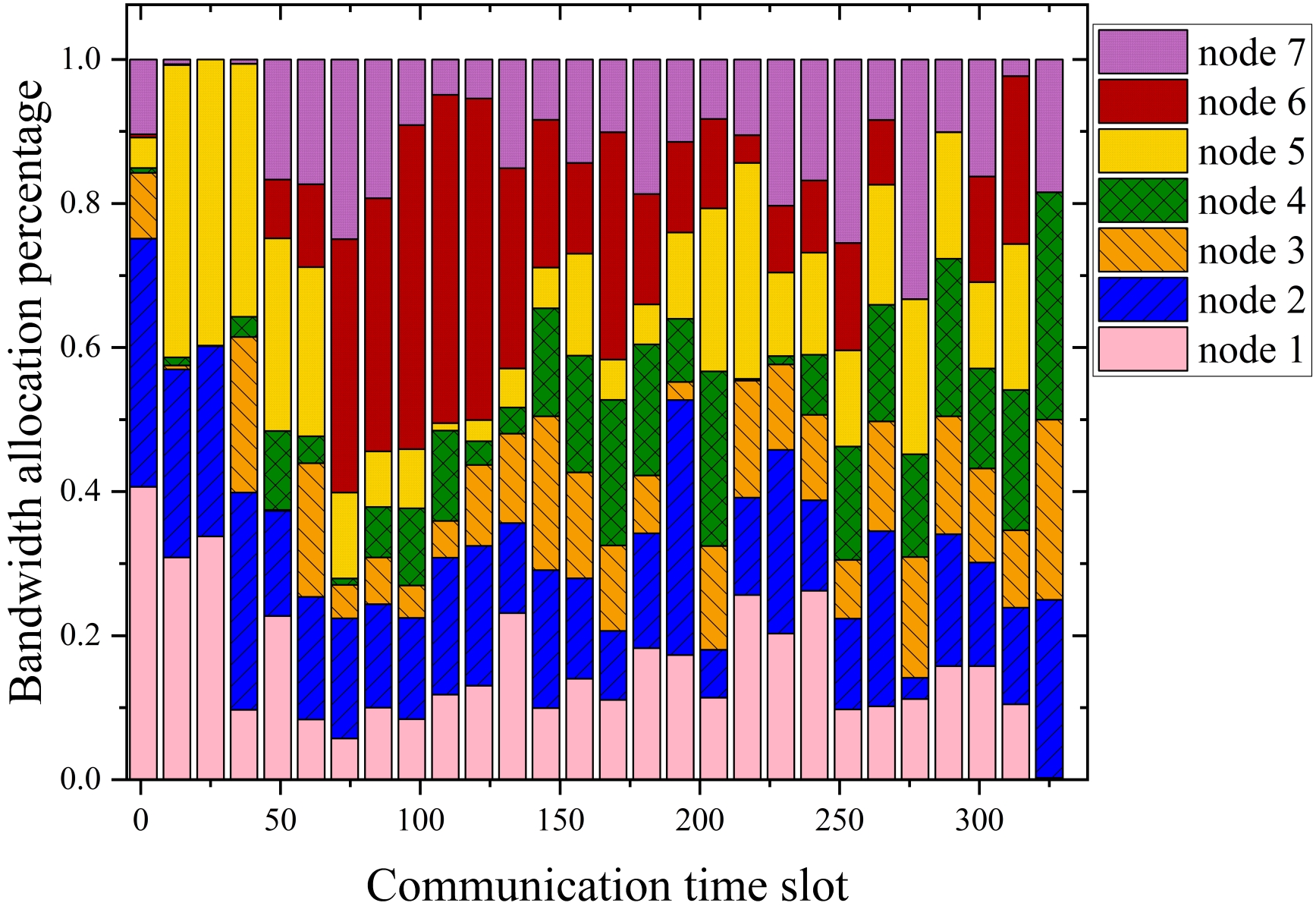}}
    \qquad
    \centering
    \subfloat[]{
    		\includegraphics[width=1.7in]{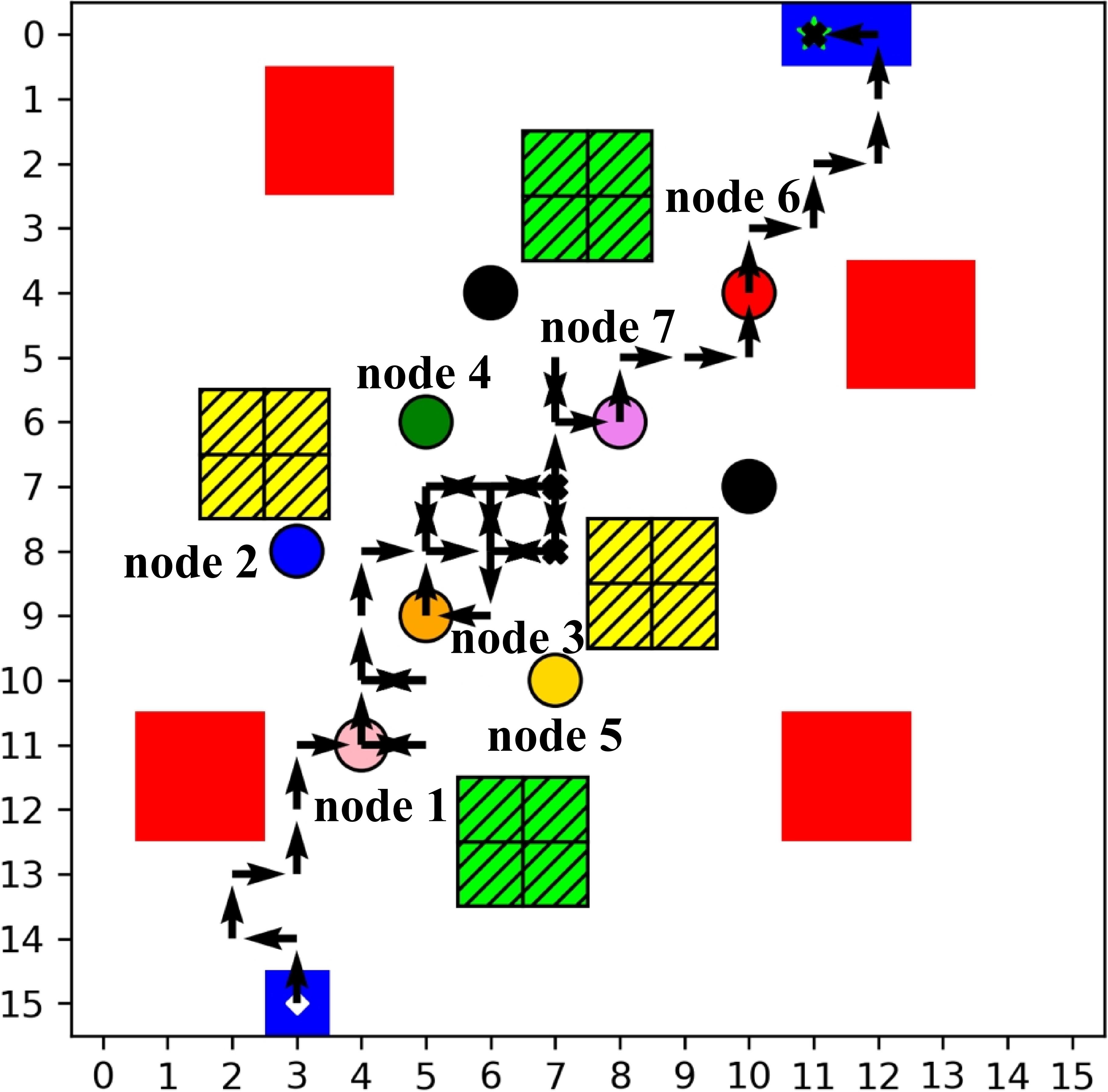}}
    \subfloat[]{
    		\includegraphics[width=2.2in]{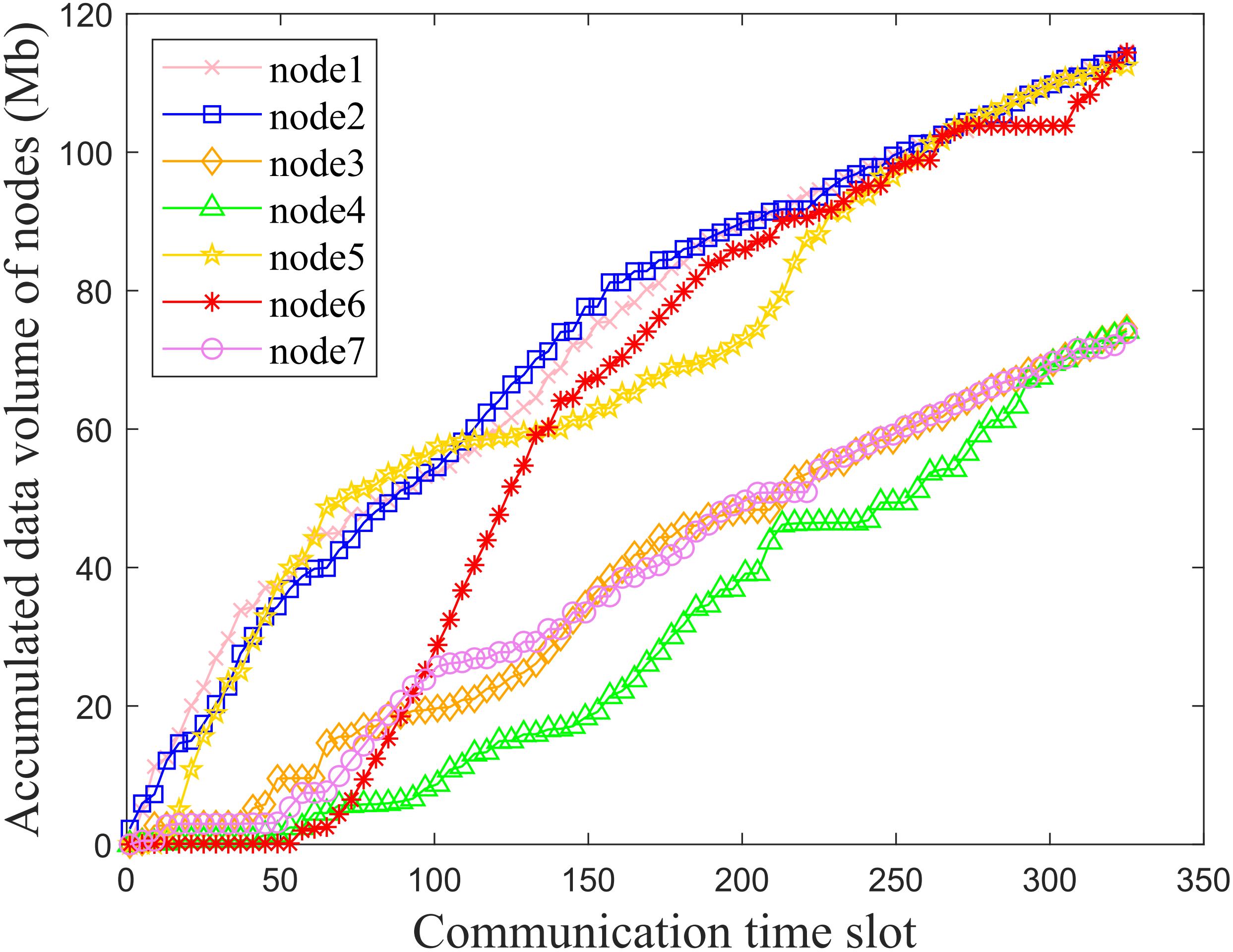}}
    \subfloat[]{
    		\includegraphics[width=2.4in]{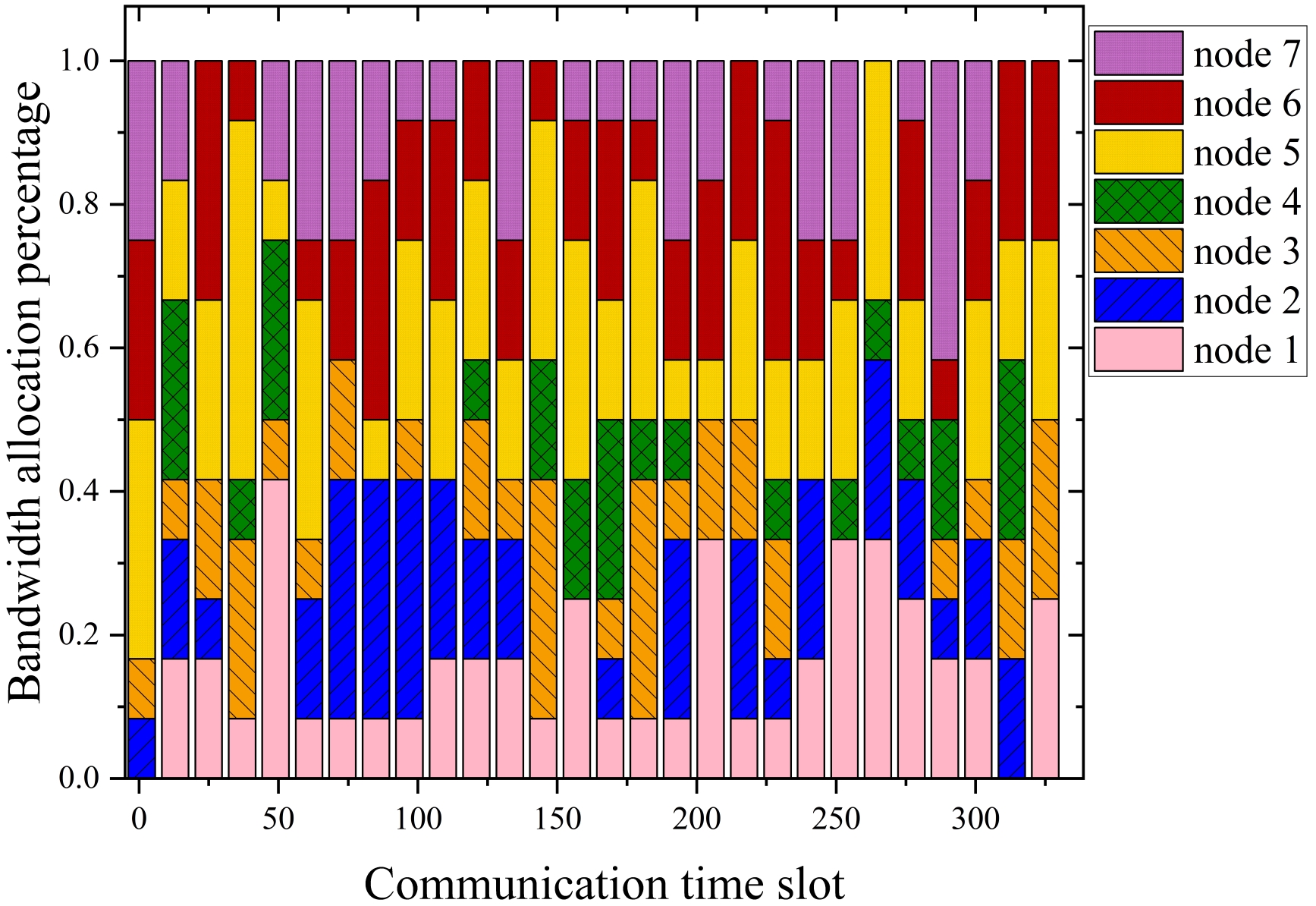}}
    \qquad
    \centering
    \subfloat[]{
    		\includegraphics[width=1.7in]{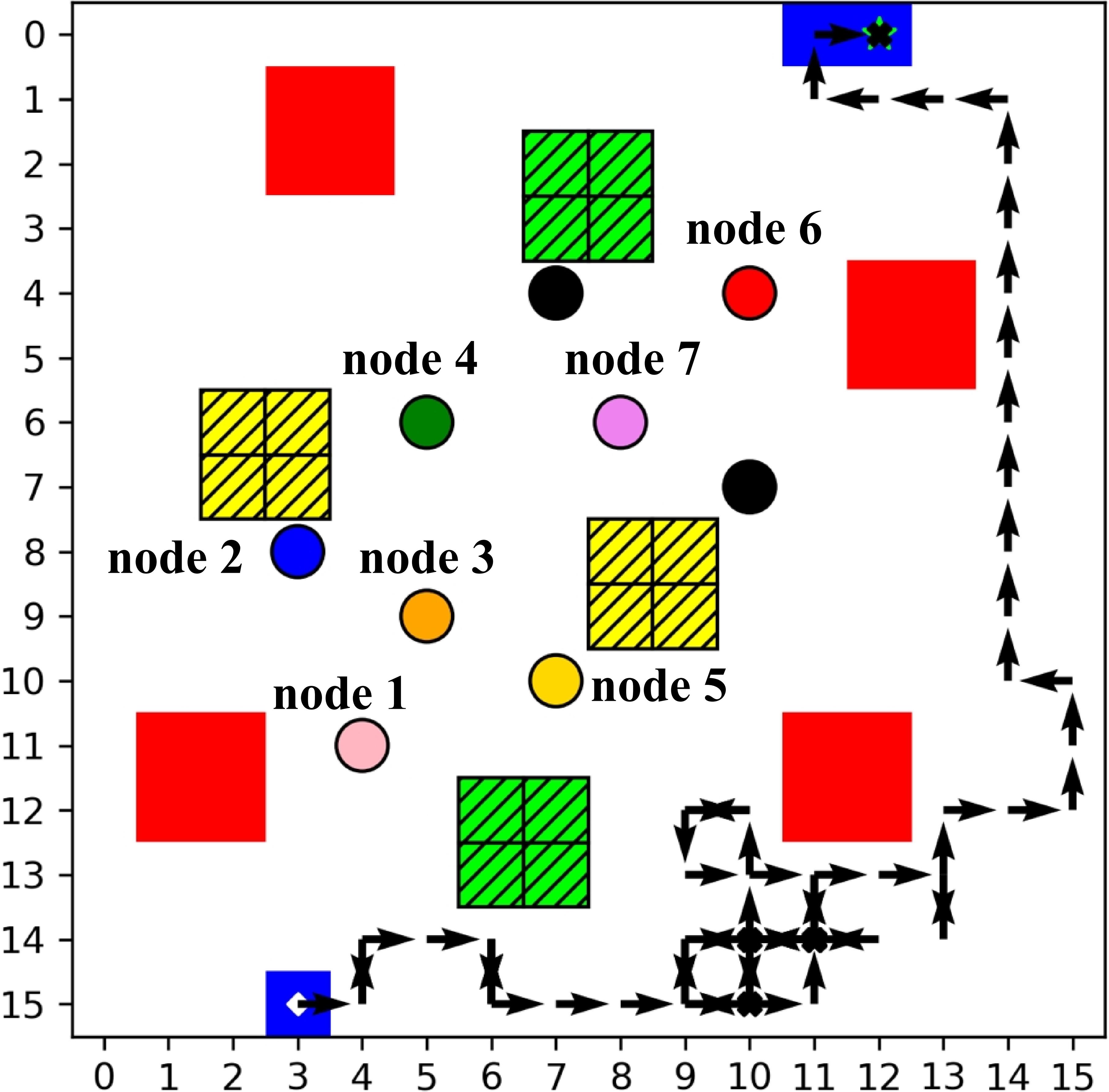}}
    \subfloat[]{
    		\includegraphics[width=2.2in]{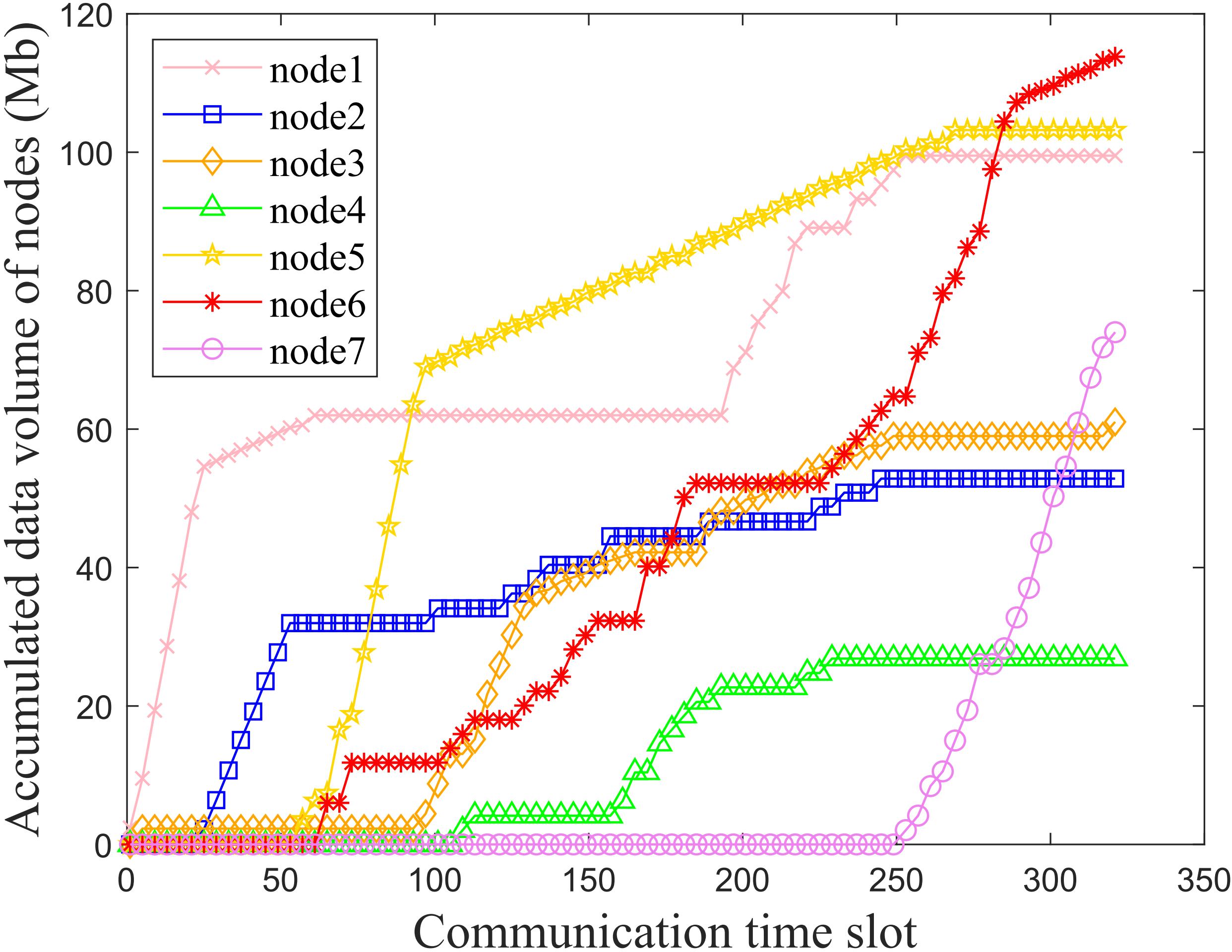}}
    \subfloat[]{
    		\includegraphics[width=2.4in]{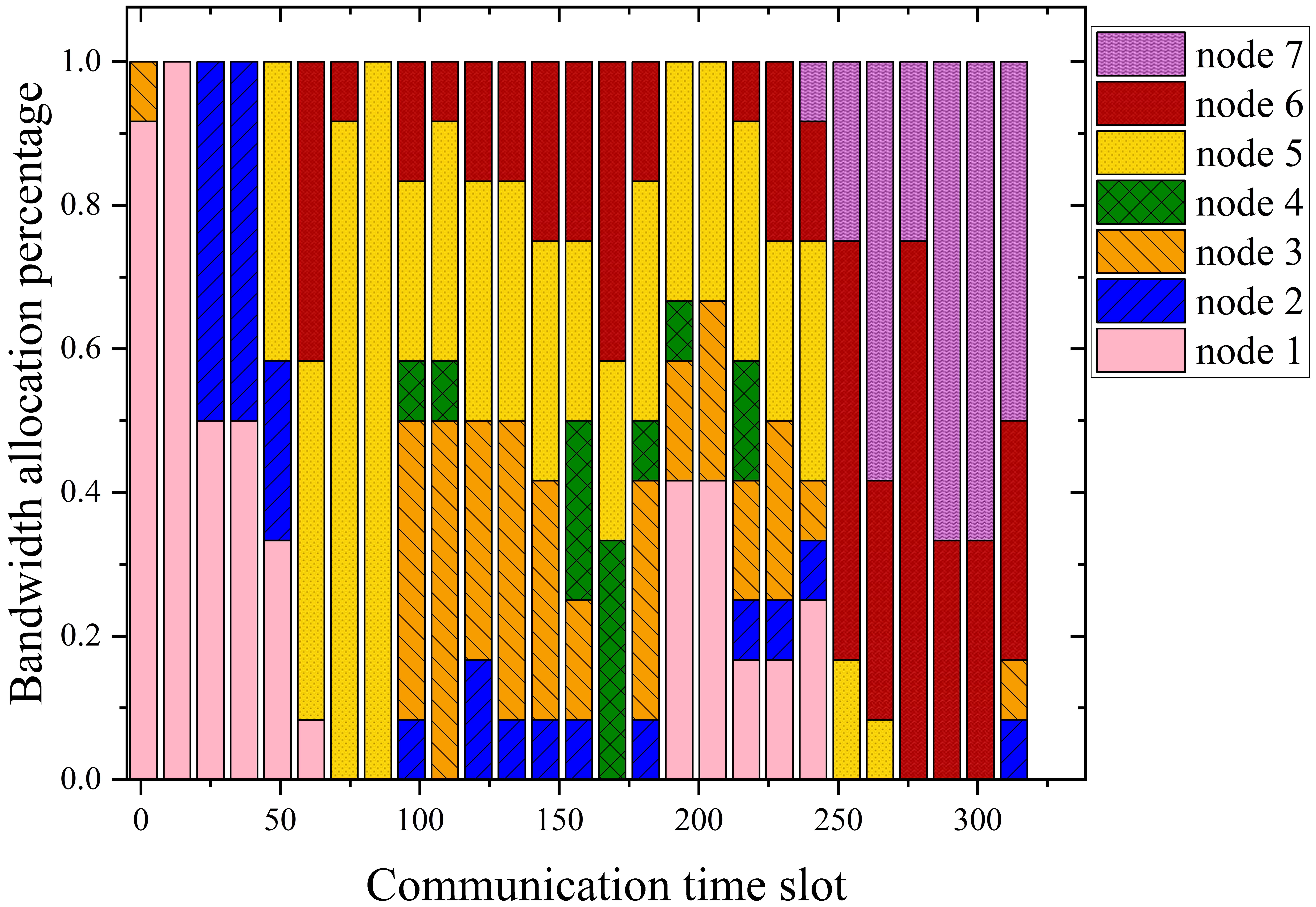}}
    \caption{The first column illustrates the trajectories of different algorithms after convergence, the second column shows the cumulative data collected by the UAV from each IoT node during each communication time slot, and the third column depicts the average bandwidth allocation ratio for every 12 communication time slots. (a), (b), and (c) are the results achieved with the TBH-DDPG algorithm; (d), (e), and (f) are the results achieved with the TBJN-DDPG algorithm; (g), (h), and (i) are the results achieved with the TBH-SAC algorithm; and (j), (k), and (l) are the results achieved with the TDMA-DDPG algorithm.}
    \label{fig7}
\end{figure*}

As shown in Fig. \ref{fig7}, the trained algorithms were tested in the given scenario. Due to the large number of communication time slots, the average bandwidth allocation ratio for every 12 communication time slots is calculated.
(1) From the figures, under both the TBH-DDPG and TBJN-DDPG algorithms, the UAV primarily collects data from IoT nodes 1, 2, 3, and 5, which are closer and have larger data volumes. As the UAV gradually moves toward the center of the map, it begins collecting data from IoT nodes 4, 6, and 7. Finally, after most of the data has been collected from all IoT nodes, the algorithms adjust bandwidth allocation to distribute excess bandwidth to farther IoT nodes, achieving balanced data collection. 
(2) The TBH-SAC algorithm, which performs slightly worse, tends to prioritize bandwidth allocation to IoT nodes with larger data volumes while paying less attention to the distance between the UAV and the IoT nodes. As a result, the cumulative data collection volume from each IoT node increases steadily, but the growth rate is faster for IoT nodes with larger data volumes. This disregard of distance weighting in bandwidth allocation slightly reduces the overall performance of TBH-SAC compared to the first two algorithms. 
(3) The TDMA-DDPG algorithm converges to a suboptimal result. The primary reason lies in its need for effective exploration during the training phase. However, since TDMA-DDPG allocates all bandwidth to the IoT nodes with the best communication link, and the data volume of IoT nodes increases over time, the penalty for data loss in the early training phase is more significant than in other algorithms. Moreover, it also faces penalties from unavoidable collisions and landing issues during exploration in the early stages. These excessive exploration penalties ultimately cause the algorithm to converge to a local optimum solution.

\subsubsection{\textbf{Impact of Data Growth Per Communication Slot on Data Loss}} 
\begin{figure}[!t]
	\centering
	\includegraphics[width=3in]{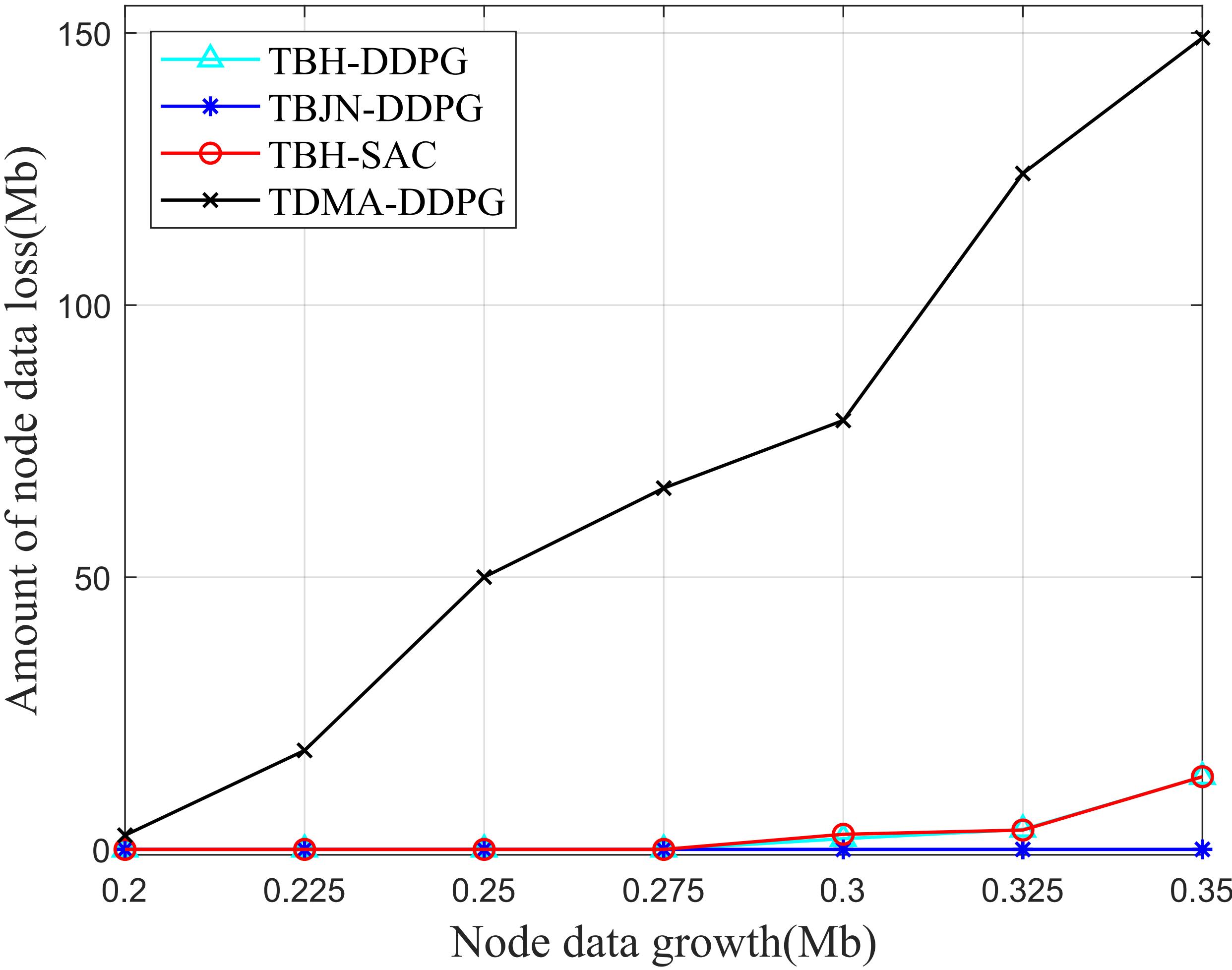}
	\caption{Impact of data growth per communication slot on data loss.}
	\label{fig8}
\end{figure}

The data volume at each IoT node increases continuously over time. When this volume exceeds the IoT node’s storage capacity, it inevitably results in data loss. As shown in Fig. \ref{fig8}, the data loss of the TDMA-DDPG algorithm increases significantly with the rise in data growth per communication slot. In contrast, other algorithms maintain relatively low data loss due to their efficient bandwidth allocation strategies, which enables a single communication slot to cover multiple IoT nodes. From the analysis in Fig. \ref{fig7}, the primary reason for the data loss is that the algorithms focus mainly on collecting data from IoT nodes located in the lower-right periphery, while collecting relatively less data from IoT nodes 2 and 4. Additionally, the TDMA-DDPG algorithm is constrained by its strategy of communicating with only one IoT node per communication slot, making it unable to effectively address the data collection demands of other IoT nodes, thereby exacerbating data loss.

\subsubsection{\textbf{Obstacle Avoidance Performance}} 
\begin{figure}[!t]
	\centering
	\includegraphics[width=3in]{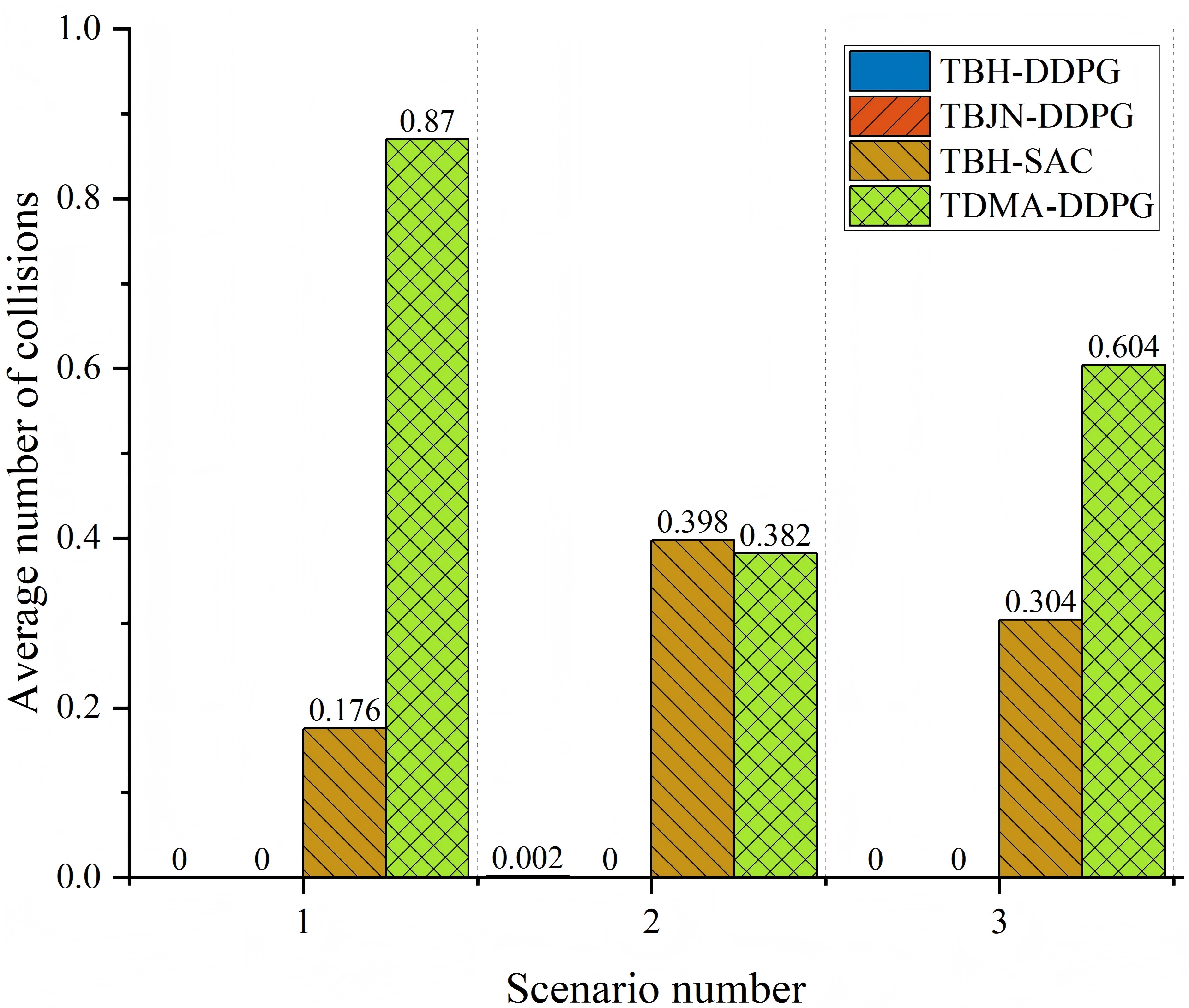}
	\caption{Average number of collisions for different algorithms in different scenarios.}
	\label{fig9}
\end{figure}

\begin{figure}[!t]
    \centering
    \subfloat[scenarios 2]{
    		\includegraphics[width=1.6in]{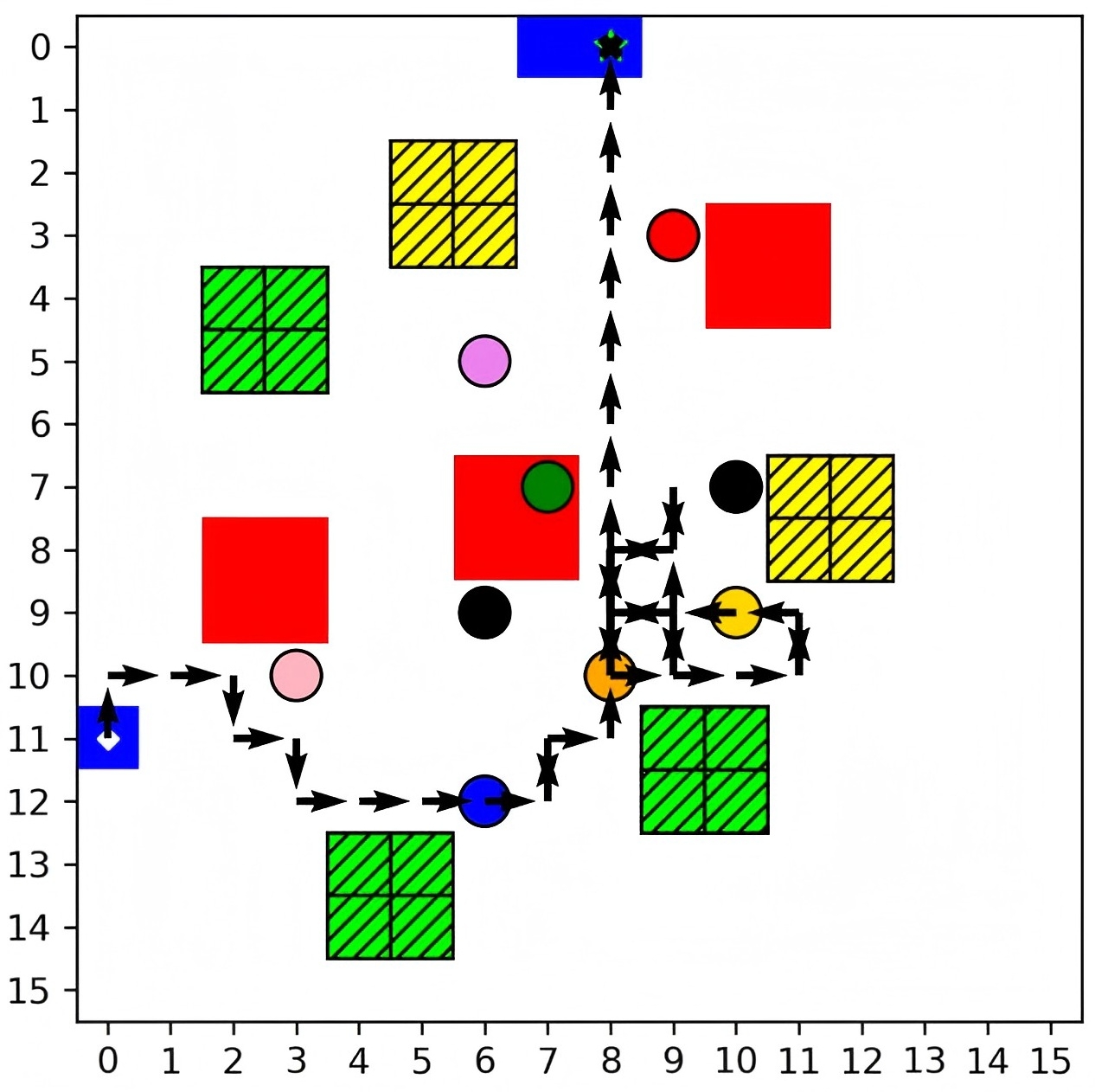}}
    \subfloat[scenarios 3]{
    		\includegraphics[width=1.6in]{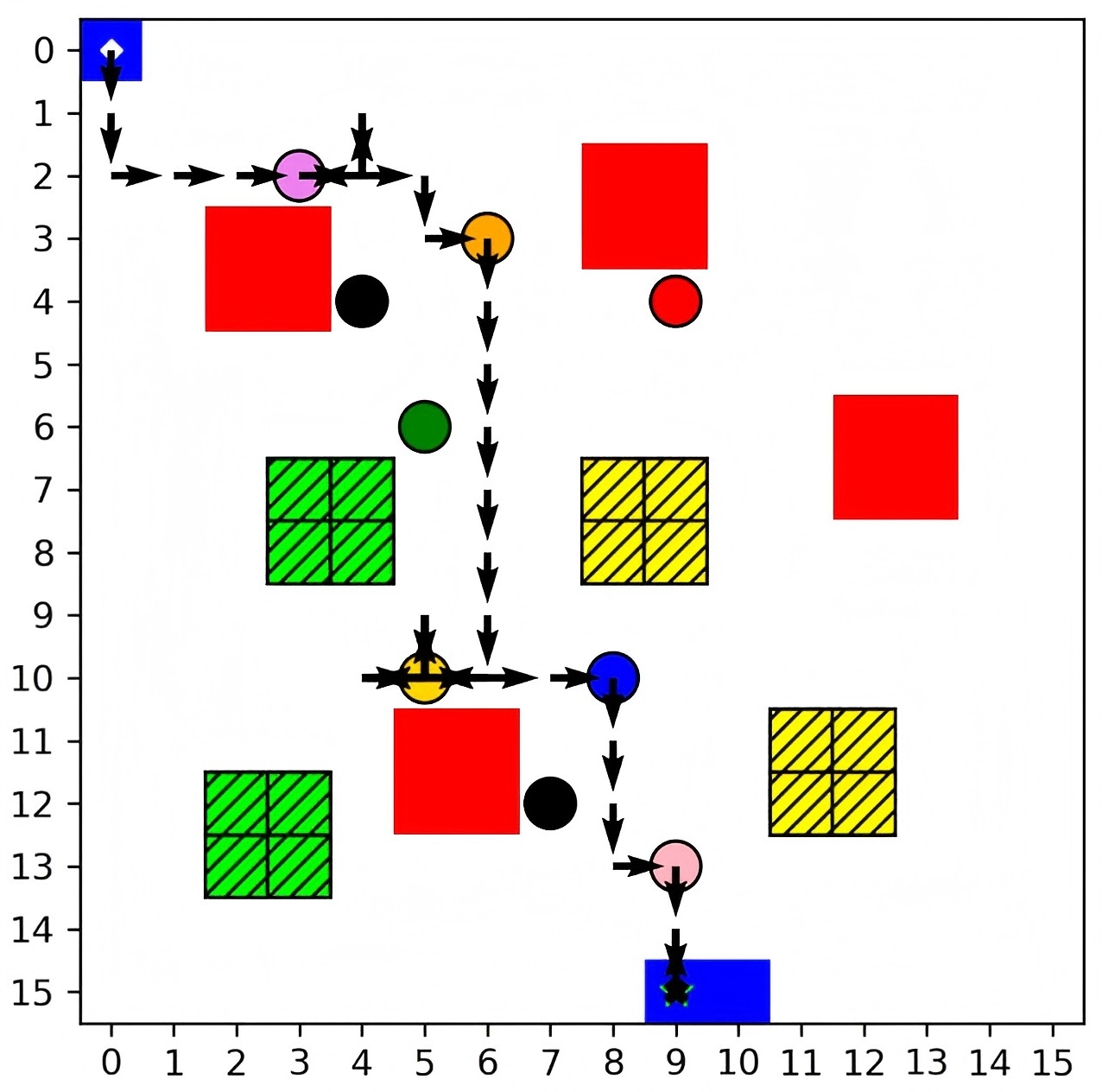}}
    \caption{UAV trajectories of TBH-DDPG algorithm in different scenarios.}
    \label{fig10}
\end{figure}

As shown in Fig. \ref{fig9}, the obstacle avoidance performance of each algorithm is tested in three different scenarios. For each scenario, 500 tests are conducted using the trained algorithms, and the average number of collisions over the 500 tests is recorded. It can be observed that the proposed TBH-DDPG algorithm and TBJN-DDPG effectively avoid collisions and successfully complete the tasks in all scenarios. Meanwhile, TBH-SAC and TDMA-DDPG experience a small number of collisions, with an average collision count of less than 1. Fig. \ref{fig10} illustrates the trajectories of TBH-DDPG in scenarios 2 and 3. The results show that TBH-DDPG not only effectively avoids obstacles and reaches the target location but also approaches data-dense regions for efficient data collection, demonstrating the robustness and adaptability of the proposed algorithm.

\subsubsection{\textbf{Impact of the Number of IoT Nodes on Total Data Collection}}
\begin{figure}[!t]
	\centering
	\includegraphics[width=3in]{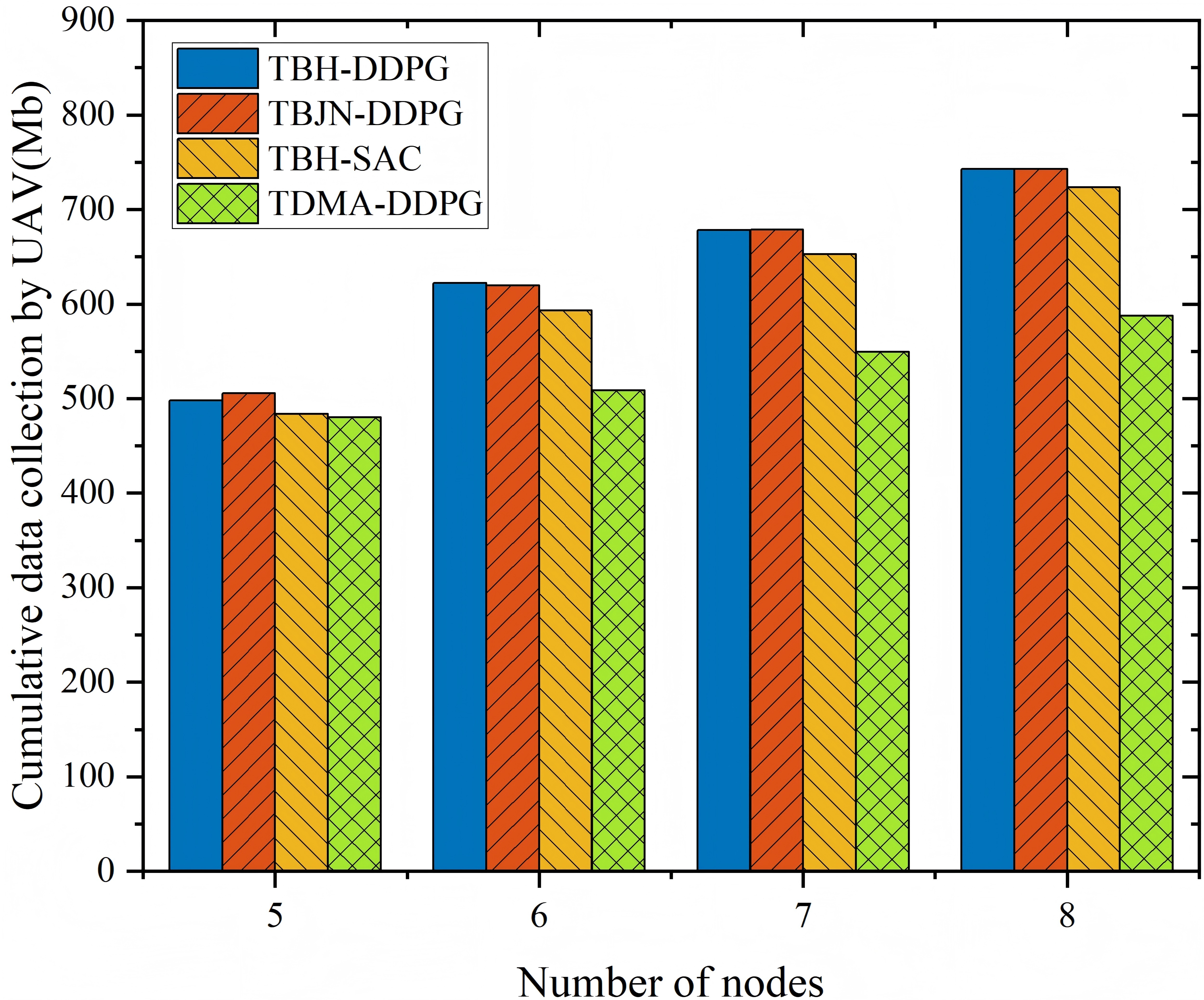}
	\caption{Impact of the number of IoT nodes on the amount of data collected by the algorithm.}
	\label{fig11}
\end{figure}

As shown in Fig. \ref{fig11}, the total data collection for all algorithms increases as the number of IoT nodes grows. The proposed TBH-DDPG algorithm outperforms the TBH-SAC and TDMA-DDPG algorithms in terms of cumulative data collection volume, and achieves performance that is nearly equivalent to that of the TBJN-DDPG algorithm. It is noteworthy that when the number of IoT nodes is relatively small, such as only five IoT nodes, the data collection performance of the TDMA-DDPG algorithm is comparable to that of algorithms employing bandwidth allocation mechanisms. Because with fewer IoT nodes, the UAV has sufficient time to traverse all IoT nodes. However, as the number of IoT nodes increases, the traversal time required also increases, causing some IoT nodes to be inadequately covered. This ultimately results in a widening gap in data collection performance between TDMA-DDPG and the algorithms with bandwidth allocation mechanisms.

\subsubsection{\textbf{Impact of Network Hidden Layers on Hierarchical and Non-Hierarchical Algorithms}}
As shown in Table \ref{tab5}, it can be observed that the non-hierarchical algorithm, TBJN-DDPG, has convergence episodes ranging from 1900 to 2400 with different hidden layer sizes, while the hierarchical algorithm, TBH-DDPG, has convergence episodes ranging from 1100 to 1600, demonstrating a significant improvement in convergence speed. Additionally, for similar rewards, the hierarchical algorithm requires only 128 hidden units in the network, while the non-hierarchical algorithm requires 1024 hidden units, resulting in a reduction in the final saved network size by approximately 6 times. Furthermore, we also provides statistics on the FLOPS required by the actor networks for both algorithms. The TBJN-DDPG algorithm has only one actor network, with a single inference during each flight period, whereas the TBH-DDPG algorithm has two actor networks: the upper-level actor infers once per flight period, and the lower-level actor network infers four times. For similar rewards, the actor network FLOPS of the hierarchical algorithm proposed is reduced by $58.05\%$. The hierarchical algorithm not only significantly improves convergence speed but also effectively reduces network size.

\begin{table*}[!t]
	\caption{Impact of network hidden layer size on the algorithm}
	\label{tab5}
	\centering
	\setlength{\tabcolsep}{2.5pt} 
	\renewcommand{\arraystretch}{1.2}
	\begin{tabular}{@{}|c|cccc|cccc|@{}}
		\toprule
		\multirow{2}{*}{\begin{tabular}[c]{@{}c@{}}\\ Hidden layer size\\of the network\end{tabular}} & \multicolumn{4}{c|}{TBJN-DDPG} & \multicolumn{4}{c|}{TBH-DDPG} \\ \cmidrule(l){2-9} 
		& \multicolumn{1}{c|}{\begin{tabular}[c]{@{}c@{}}Average\\ reward\end{tabular}} & \multicolumn{1}{c|}{\begin{tabular}[c]{@{}c@{}}Converged\\ episode range\end{tabular}} & \multicolumn{1}{c|}{\begin{tabular}[c]{@{}c@{}}Saved network\\ size (bit)\end{tabular}} & \begin{tabular}[c]{@{}c@{}}FLOPS of\\ the actor networks\end{tabular} & \multicolumn{1}{c|}{\begin{tabular}[c]{@{}c@{}}Average\\ reward\end{tabular}} & \multicolumn{1}{c|}{\begin{tabular}[c]{@{}c@{}}Converged\\ episode range\end{tabular}} & \multicolumn{1}{c|}{\begin{tabular}[c]{@{}c@{}}Saved network\\ size (bit)\end{tabular}} & \begin{tabular}[c]{@{}c@{}}FLOPS of\\ the actor networks\end{tabular} \\ \midrule
		1024 & \multicolumn{1}{c|}{\textbf{1010.45}} & \multicolumn{1}{c|}{\textbf{2100-2400}} & \multicolumn{1}{c|}{\textbf{91,856,566}} & \textbf{2865152} & \multicolumn{1}{c|}{1020.44} & \multicolumn{1}{c|}{1300-1600} & \multicolumn{1}{c|}{181,991,448} & 14202880 \\ \midrule
		512 & \multicolumn{1}{c|}{970.31} & \multicolumn{1}{c|}{2100-2400} & \multicolumn{1}{c|}{37,584,566} & 1170432 & \multicolumn{1}{c|}{1018.45} & \multicolumn{1}{c|}{1200-1500} & \multicolumn{1}{c|}{74,274,456} & 5790720 \\ \midrule
		256 & \multicolumn{1}{c|}{916.41} & \multicolumn{1}{c|}{1900-2200} & \multicolumn{1}{c|}{16,740,022} & 519680 & \multicolumn{1}{c|}{1013.79} & \multicolumn{1}{c|}{1200-1500} & \multicolumn{1}{c|}{32,999,064} & 2567680 \\ \midrule
		128 & \multicolumn{1}{c|}{846.57} & \multicolumn{1}{c|}{2000-2300} & \multicolumn{1}{c|}{7,890,486} & 243456 & \multicolumn{1}{c|}{\textbf{1009.91}} & \multicolumn{1}{c|}{\textbf{1100-1400}} & \multicolumn{1}{c|}{\textbf{15,506,968}} & \textbf{1201920} \\ \bottomrule
	\end{tabular}
\end{table*}

\subsubsection{\textbf{Impact of Key Hyperparameters on the Proposed Algorithm}}
As shown in Table \ref{tab_ablation}, we evaluate the impact of the soft update parameter $\tau$, exploration noise, and discount factor $\gamma$ on the proposed algorithm. The proposed setting achieves an average reward of 1009.91 and converges within 1100-1400 episodes, showing a good trade-off between convergence speed and final performance. For $\tau$, a larger value reduces the average reward, while a smaller value slightly improves the reward but significantly delays convergence. For exploration noise, insufficient noise limits exploration and leads to poor performance, whereas excessive noise delays convergence. For $\gamma$, a smaller value slightly reduces the final reward, while a larger value requires more episodes to converge. Overall, the proposed hyperparameter setting provides a stable and efficient configuration for the TBH-DDPG algorithm.

\begin{table}[!t]
	\caption{Impact of Key Hyperparameters on the Proposed Algorithm}
	\label{tab_ablation}
	\centering
	\footnotesize
	\setlength{\tabcolsep}{2.0pt}
	\renewcommand{\arraystretch}{1.15}
	\begin{tabular}{@{}|p{1.35cm}|c|c|c|c|@{}}
		\toprule
		Parameter & Upper level & Lower level & Avg. reward & Conv. range \\ 
		\midrule
		Proposed & Proposed & Proposed & 1009.91 & 1100--1400 \\ 
		\midrule
		\multirow{2}{1.4cm}{Soft update parameter $\tau$} 
		& 0.01 & 0.0001 & 944.35 & 1200--1500 \\ 
		\cmidrule(l){2-5}
		& 0.001 & 0.000001 & 1014.75 & 1900--2100 \\ 
		\midrule
		\multirow{2}{1.4cm}{Exploration noise} 
		& 1 & 0.5 & 900.67 & 1600--1900 \\ 
		\cmidrule(l){2-5}
		& 5 & 2 & 1015.16 & 1600--1900 \\ 
		\midrule
		\multirow{2}{1.4cm}{Discount factor $\gamma$} 
		& 0.9 & 0.9 & 1003.91 & 1100--1400 \\ 
		\cmidrule(l){2-5}
		& 0.999 & 0.999 & 1010.20 & 1400--1700 \\ 
		\bottomrule
	\end{tabular}
\end{table}

\subsubsection{\textbf{Impact of Imperfect CSI and Residual Interference on the Proposed Algorithm}}
\begin{figure}[!t]
	\centering
	\subfloat[Impact of CSI uncertainty]{
		\includegraphics[width=1.6in]{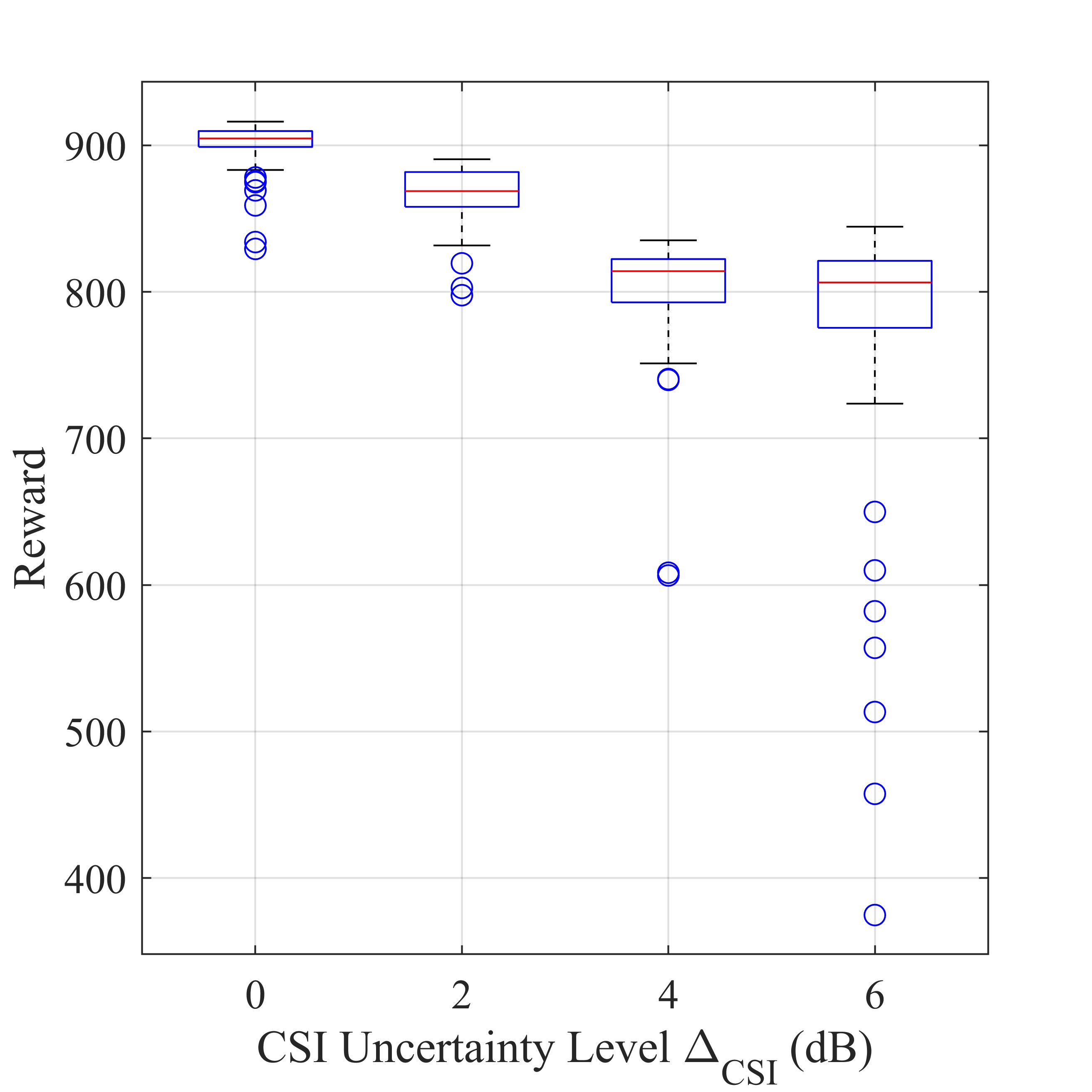}}
	\subfloat[Impact of residual interference]{
		\includegraphics[width=1.6in]{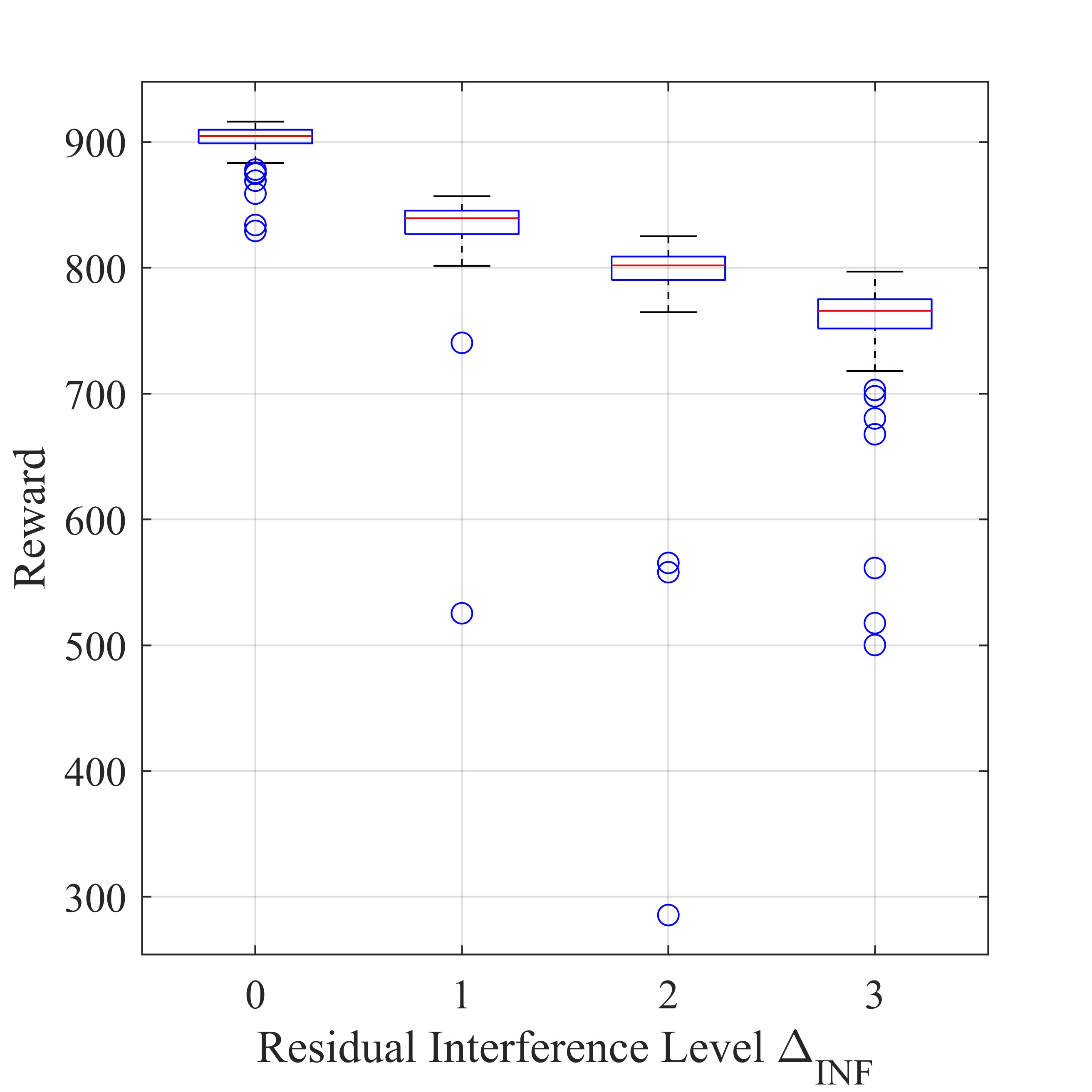}}
	\caption{Robustness evaluation of the proposed algorithm under imperfect CSI and residual interference.}
	\label{fig12}
\end{figure}

To further evaluate the robustness of the proposed algorithm under non-ideal channel conditions, the total available bandwidth is reduced from $0.5$ MHz to $0.35$ MHz to construct a more resource-constrained scenario, where the effects of CSI uncertainty and residual interference can be more clearly observed. Fig. \ref{fig12} presents the box plots of rewards collected from 100 consecutive episodes in the stable convergence stage. As shown in Fig. \ref{fig12}(a), increasing $\Delta_{\mathrm{CSI}}$ from $0$ to $6$ dB gradually lowers the reward distribution and enlarges its fluctuation, indicating that channel-gain perturbations degrade decision quality. Similarly, Fig. \ref{fig12}(b) shows that larger $\Delta_{\mathrm{INF}}$ leads to a lower median reward and more low-reward samples, since residual interference reduces the effective SINR. Nevertheless, the proposed algorithm remains stably convergent under both non-ideal conditions, demonstrating its robustness to imperfect CSI and residual interference.

\section{Conclusion}
We primarily focused on UAV trajectory planning and bandwidth allocation in the context of data collection for urban environments within low-altitude IoT systems. The TBH‑DDPG algorithm was proposed to enable a UAV to maximize data collection from distributed IoT nodes, with the upper layer optimizing the flight trajectory and the lower layer optimizing the bandwidth allocation. The experimental results showed that, compared to the baseline algorithms, the proposed algorithm improved convergence speed by $44.44\%$, reduced computational complexity by $58.05\%$, and effectively prevented data loss. 

Moreover, across various scenarios, the proposed algorithm not only enabled the UAV to effectively avoid collisions but also ensured it continues to collect data efficiently as the number of IoT nodes increases. However, the algorithm had limitations, such as complex hyperparameter tuning tied to scenario-specific conditions. Future work will deploy the algorithm on real hardware to evaluate its performance and stability. We also plan to explore its application in larger-scale multi-UAV collaborative scenarios, emphasizing UAV speed and direction continuity to handle dynamic tasks. HDRL enables agents to train independently for specific tasks, allowing UAV swarms to collaborate efficiently, reducing task complexity and alleviating computational demands on individual UAVs.

\section{Acknowledgments}
\label{sec11}
This work was supported in part by the Postgraduate Research and Practice Innovation Program of NUAA (No. xcxjh20240406)

\bibliography{reference_clean}
\bibliographystyle{iet}

\end{document}